\documentclass[aps,pra,twocolumn,showpacs,showkeys]{revtex4}
\usepackage{graphicx}
\usepackage{dcolumn}
\usepackage{amssymb} 
 
\usepackage{amsfonts} 
\usepackage{amssymb} 
\usepackage{amsmath} 
\usepackage{color}
\usepackage{hyperref}
 
\newcommand{\ar}{\arrowvert}

\newcommand{\cd}{\! \cdot \!} 
\newcommand{\be}{\begin{equation}} 
\newcommand{\ee}{\end{equation}} 
\newcommand{\ba}{\begin{eqnarray}} 
\newcommand{\ea}{\end{eqnarray}}
\newcommand{\x}{x} 
\newcommand{\dx}{\dot{x}}  
\newcommand{\ddx}{\ddot{x}}
\newcommand{\dddx}{\dddot{x}}
\newcommand{\eref}[1]{~(\ref{#1})}

\begin{document} 
\title{Radiation reaction on a classical charged particle: \\
a modified form of the equation of motion}
\author{  Guillermo Garc\'{\i}a Alcaine and Felipe J. Llanes-Estrada}
\email{fllanes@fis.ucm.es} 
\affiliation{ 
Departamento de F\'{\i}sica Te\'orica I,  Universidad 
Complutense, 28040 Madrid, Spain. 
} 
 
\date{\today}

\begin{abstract}  
We present and numerically solve a modified form of the equation of motion for a charged
particle under the influence of an external force, taking into account the radiation reaction.
This covariant equation is integrodifferential, as Dirac-R\"ohrlich's, but has several technical improvements.
First, the equation has the form of the second Newton law, with acceleration isolated on the left hand side, and the force depending only on positions and velocities: thus, the equation is linear in the highest derivative.
Second, the total four-force is by construction perpendicular to the four-velocity. And third, if the external force vanishes for all future times, the total force and the acceleration automatically vanish at present time.
We show the advantages of this equation by solving it numerically for several examples of external force.
\end{abstract} 
%
\maketitle

\section{Introduction} \label{sec:intro}

Accelerated particles carrying electric charge are a source of electromagnetic radiation. The momentum carried away by the radiation field affects the particles's classical motion imparting a recoil force traditionally known as ``radiation reaction''. It is somewhat frustrating that a century and a half after the first attempts, ``a completely satisfactory classical treatment of the reactive effects of radiation does not exist''~\cite{Jackson3}.
Other treatments employing QED have been proposed, for example
~\cite{Krivitskii,Johnson}.
But nevertheless we feel that classical treatments are still interesting as attested by the continuing theoretical efforts~\cite{Harvey:2010ns,Martin:2008nr,Spohn:1999uf,Matolcsi}.

Most physicists are familiar with 
the Abraham and Lorentz~\cite{Jackson} equation 
of motion~\footnote{Dots indicate laboratory time derivatives when acting on a (boldfaced) three-dimensional vector, but derivatives with respect to invariant interval $s$ (with dimension of proper time multiplied by $c$) of the moving particle when acting on a four-vector.}
\be \label{ALeq}
m{\bf \ddx}= {\bf f}_{ext}({\bf \x},{\bf \dx}) + \frac{2}{3} \frac{e^2}{c^3} {\bf \dddx}
\ee
that incorporates a radiation reaction force $\frac{2}{3} \frac{e^2}{ c^3} {\bf \dddx}$ into Newton's second law, in addition to any external forces ${\bf f}_{ext}$ accelerating the particle. 
(Usually $m$ and $e$ will be the mass and charge of the electron, but the equation will be valid for any pointlike charged system.) An alternative to Eq.~(\ref{ALeq}), the Landau-Lifshitz equation, has been carefully compared in a very  clear way in~\cite{Griffiths}. 
Although the Landau-Lifschitz formulation has some advantages, stating that it is the correct one seems an overstatement, as  remarked in~\cite{Griffiths}. The results in this work are based on the classical Abraham-Lorentz equation and its generalizations.

A well known one is the relativistic four-vector form, known as Lorentz-Dirac equation
\be \label{LorentzDirac}
m\ddx^\mu(s) = f^\mu_{ext}(s) + \frac{2}{3c^2}e^2
\left( \dddx^\mu(s)+ (\ddx(s))^2\dx^\mu(s) \right)
\ee
whose derivation can be found in standard textbooks~\cite{Barut,Rohrlich}.
A few remarks will however be carried onto this work later on in section~\ref{sec:demo}.

This equation has traditionally being recognized as affected by the problem of self-accelerated solutions, discussed later in section~\ref{sec:demo} and subsection~\ref{subsec:causality}. Therefore, it is often replaced by an integral equation, the Dirac-R\"ohrlich equation~\cite{RohrlichAP,Plass:1961zz,Klepikov}, 
\be \label{Rohrlich}
\ddx^\mu(s)= \int_s^\infty ds' e^{(s-s')/L}  
\left(\frac{f_{ext}^\mu(s')}{m L} +(\ddx(s'))^2\dx^\mu(s') \right) \ . 
\ee 
This is an integro-differential equation for the acceleration, featuring a characteristic length,  
$L\equiv 2 e^2/(3m c^2)$ that for the electron is called the ``electron relaxation length'', $L_e=1.876$fm.
The self accelerated solutions of Eq.~(\ref{LorentzDirac}) are absent by construction (the formal integration is carried out with the condition that the acceleration does not increase exponentially as $e^{s/L}$ or faster at infinity). 

We present in this work a novel equation, equivalent to~Eq.(\ref{Rohrlich}), but that has several important technical improvements. This is
\be \label{NewRadRecEq}
\boxed{
\ddx^\mu(s) = \frac{1}{mL}\dx_\nu(s) \int_s^\infty ds' e^{(s-s')/L}
\left(f^\mu_{ext}\dx^\nu-f^\nu_{ext}\dx^\mu \right)(s')
} 
\ee

Equation\eref{Rohrlich} is indeed known to have several drawbacks. 
The first one is the apparent lack of causality as the acceleration depends on the force at future times, but this ``advanced'' formulation is not in itself a real difficulty as we will show in section~\ref{sec:Newton}. The second and most serious drawback is the phenomenon of ``preacceleration'', at scales of order $L$ where the particle accelerates before the external force begins.
In this respect our equation will not improve the existing situation.
Nevertheless, our alternative equation\eref{NewRadRecEq} is superior
in several important technical counts, that make its numerical implementation and solution much simpler. These are
\begin{enumerate}
\item Our equation is linear in the highest (second) derivative, while Eq.~(\ref{Rohrlich}) is quadratic in it.
\item The acceleration $\ddx$ appears in both left and right hand side of 
Eq.~(\ref{Rohrlich}), while in our proposed equation Eq.~(\ref{NewRadRecEq}) the right hand side depends only on four-position and four-velocity, but the four-acceleration is formally solved for. 
\item If the external force vanishes for all times, the new equation~(\ref{NewRadRecEq}) yields zero acceleration automatically. 
This happens also in Eq.~(\ref{Rohrlich}) although it is not as obvious.
\item The right hand side of Eq.~(\ref{Rohrlich}) is orthogonal to the four velocity, 
but this is not evident.  In our equation~(\ref{NewRadRecEq}), 
orthogonality follows trivially from the antisymmetry of the integrand under $\mu\leftrightarrow \nu$. 
This orthogonality guarantees $\dx_\mu\ddx^\mu=0$ and thus $\dx^2={\rm constant}$ at every step on an iterative solution without relying on cancellations not always obvious.
\end{enumerate}

We will derive, analyze and numerically solve equation~(\ref{NewRadRecEq}) for several simple cases of interest.
Our numerical methods are described in section~\ref{sec:nummethod} and the actual computations are reported in section~\ref{sec:Numeric}. Section~\ref{sec:conclusions} wraps up the discussion and summarizes our findings.

%
%

\section{Derivation of the new equation} \label{sec:demo}
\subsection{Self-accelerated solutions}

The Lorentz-Dirac equation~(\ref{LorentzDirac}) features two radiation-related self-forces. One is $mL(\ddx(s))^2\dx^\mu(s)$, that accounts for the power radiated (see below eq.~(\ref{radpower}) ) in an irreversible manner. 
The other, the Schott term $mL\dddx^\mu(s)$,  also present in the relativistic generalization of the Abraham-Lorentz equation (see the derivation in~\cite{Barut}), represents a reversible positive or negative  transfer of four-momentum between the charged particle and its near field. 
This structure entails that setting $f_{ext}^\mu(s)=0$ for all times is not sufficient to guarantee $\ddx=0$ in Eq.\eref{LorentzDirac}.
Moreover, the Schott term brings about self-accelerated solutions with exponentially diverging accelerations.

Indeed, the ``free'' Dirac-Lorentz equation
\be
\ddx^\mu(s) = L \left( \dddx^\mu(s)+(\ddx)^2\dx^\mu\right) 
\ee
can be multiplied by $\ddx_\mu$, and using the orthogonality between four-velocity and four-acceleration $\dx\cd\ddx=0$ the radiated power term vanishes,
leaving
\be
L \ddx\cd \dddx= \ddx^2
\ee
that can be integrated to yield a family of solutions characterized by an arbitrary real constant $b$
\be \label{selfaccelerated}
\ddx^2= -b^2e^{\frac{2s}{L}} \ .
\ee
This solution is self-accelerated, in the sense that for large proper time $s$ the acceleration grows exponentially as $e^{s/L}$, a clearly unphysical behavior.
The integration constant $b$ happens to be the 
component of the three-dimensional acceleration at $s=0$ that is parallel to the velocity, \\
${\bf a}(0)\ \cd\ {\bf v}(0)=b\ar{\bf v}(0)\ar$, as can be seen employing the 
reduction in equation~(\ref{acceleration}) below.

Thus, customarily the Dirac-Lorentz equation is complemented by a physical boundary condition requiring the acceleration to vanish at infinite time,
$\lim_{s\to\infty} \ar {\bf a}\ar\to 0$ (free particle), or a less restrictive one asking that the product 
\be \label{boundary1}
\lim_{s\to\infty} e^{-s/L}\ar {\bf a}\ar\to 0
\ee 
so that the aceleration does not grow faster than $e^{s/L}$, thus getting rid of the runaway solutions of the free case.
Any of these boundary conditions at infinity is sufficient to eliminate the self-accelerated solutions and yields the expected behavior
\be \label{babyNewton1}
f^\mu_{ext}(s)=0\ \ \forall s\ \  \Rightarrow\ \ \ddx^\mu(s)=0\ \ \forall s\ .
\ee

\subsection{Construction of the Dirac-Rohrlich equation}

Let us briefly remind the reader how the traditional equation of motion 
Eq.~(\ref{Rohrlich}) is obtained. Start by multiplying Eq.~(\ref{LorentzDirac}) by an integrating factor $e^{-s/L}$. The Dirac-Lorentz equation then reads
\be
e^{-s/L}\left( \frac{\ddx^\mu}{L}-\dddx^\mu\right)=
e^{-s/L} \left(\frac{f^\mu_{ext}}{mL}+(\ddx)^2\dx^\mu \right)\ .
\ee
The left hand side can be then written as a total derivative
\be \label{restartingpoint}
 -\frac{d}{ds}\left(e^{-s/L}\ddx^\mu\right) = e^{-s/L} \left(\frac{f^\mu_{ext}}{mL}+(\ddx)^2\dx^\mu \right)\ .
\ee
This can be integrated to yield an integral equation where the integration constant sets the acceleration at initial proper time $s_i$ (since the starting point was the third order Lorentz-Dirac equation), yielding
\ba \label{eqintermedia}
 e^{-s/L} \ddx^\mu(s) &= &
e^{-s_i/L}\ddx^\mu(s_i)  \\ \nonumber  & &
- \int_{s_i}^s ds' e^{-s'/L} \left(\frac{f^\mu_{ext}}{mL}+(\ddx)^2\dx^\mu \right)(s') \ .
\ea
In this expression we already see that the right-hand side builds up a four-force that is non-local in time. Imposing now the four conditions to the acceleration
\be \label{boundary2}
\lim_{s\to \infty}\left( e^{-s/L}\ddx^\mu(s)\right) =0
\ee
that eliminate the self-accelerated solutions of the {\emph{free}}
Lorentz-Dirac equation, we can evaluate Eq.~(\ref{eqintermedia}) at $s=\infty$ to read
\be
e^{-s_i/L} \ddx^\mu(s_i) =  
\int_{s_i}^\infty ds' e^{-s'/L} \left(\frac{f^\mu_{ext}}{mL}+(\ddx)^2\dx^\mu \right)(s') 
\ee
that, upon substitution in Eq.~(\ref{eqintermedia}) and combining the two integrals, results in Eq.~(\ref{Rohrlich}).
The self-accelerated solutions of the free equation satisfying Eq.\eref{selfaccelerated} have been eliminated
\footnote{This can be checked by noting that the left hand side of  Eq.~(\ref{Rohrlich}), the four-acceleration, is a spacelike four-vector, while the right hand side in the absence of the external force must be timelike, so that the equality can only be satisfied by $\ddx^\mu=0$.}. This however is only certain when the external force is identically zero, for all times.

Several authors have attempted to provide more satisfactory equations in one or another respect. For example, Kazinski and Shipulya~\cite{Kazinski:2010ce} have isolated the acceleration on the left hand side of the equation in a simplified two-dimensional problem with constant electromagnetic field. We find this an attractive way of proceeding and show here the solution to this problem for arbitrary dimension and arbitrary electromagnetic fields.

\subsection{New treatment}\label{Newtreatment}

To obtain equation\eref{NewRadRecEq} we multiply Eq.\eref{restartingpoint} by $\dx^\nu(s)$, which turns the vector family of equations with an index $\mu$ into a tensor family with $\mu$, $\nu$, and we antisymmetrize in the two indices $\mu$, $\nu$. 
This manipulation has the merit of cancelling the non-linear radiation term 

Barut~\cite{Barut} employed the same antisymmetrization in the absence of external forces to identify and then eliminate self-accelerating unphysical solutions, but did not extend it to the general case with external forces as we do here.

We are left with a differential equation equivalent to the Lorentz-Dirac equation but where the non-linearity in the highest derivative has been traded for a higher-rank tensor structure
\ba \label{intermediateNew}
\dx^\nu(s)\frac{d}{ds}\left( e^{-s/L}\ddx^\mu(s) \right) - (\mu\leftrightarrow\nu) =
\nonumber \\
-\frac{e^{-s/L}}{mL}\left( \dx^\nu f^\mu_{ext}- \dx^\mu f^\nu_{ext}
\right)\ .
\ea
The left hand side of this equation is an exact derivative. Let us shorten the notation defining two auxiliary antisymmetric tensors,
\ba \label{KandC}
C^{\mu\nu}& \equiv & \ddx^\mu\dx^\nu -\ddx^\nu\dx^\mu\\ \label{onlyK}
K^{\mu\nu}& \equiv & f^\mu_{ext} \dx^\nu - f^\nu_{ext} \dx^\mu
\ea
that, substituted in Eq.\eref{intermediateNew} turn it into~\footnote{
This in itself is an interesting equation among two exterior products. The right hand side is similar to the familiar mechanical torque by substituting position by velocity, and it may be called ``velocity-torque''. Likewise, the tensor in the left hand side is a sort of angular momentum with higher derivatives. 
}
\be
\frac{d}{ds} \left( e^{-s/L}C^{\mu\nu}(s) \right) = -\frac{e^{-s/L}}{mL} K^{\mu\nu}(s)\ .
\ee
A formal integration leads to an expression analogous to Eq.\eref{eqintermedia}, in terms of the initial condition for $C^{\mu\nu}(s_i)$. 

We now restrict ourselves to forces that vanish when $s\to \infty$ and solutions
satisfying the asymptotic condition (that eliminates self-accelerated solutions of the free equation of motion),
\be \label{boundary3}
\lim_{s\to \infty} \left( e^{-s/L}C^{\mu\nu}(s)\right)=0
\ee
so we obtain
\be \label{casifinal}
e^{-s/L}C^{\mu\nu}(s) = \frac{1}{mL}\int_s^\infty ds' e^{-s'/L} K^{\mu\nu}(s') \ .
\ee
To return this equation to one with the acceleration solved for, as in Newton's second law, we employ the property $\dx_\nu C^{\mu\nu}=\ddx^\mu$ that follows from $\dx^2=1$. Multiplying Eq.\eref{casifinal} by $\dx^\nu$ and applying this property, 
\be \label{penultima}
\ddx^\mu(s) = \frac{\dx_\nu(s)}{mL} \int_s^\infty ds' e^{(s-s')/L}K^{\mu\nu}(s')
\ee
which is Eq.\eref{NewRadRecEq}, and completes the demonstration.

Finally we note the form that the equation takes if the external force is of electromagnetic nature,
\ba
\ddx^\mu(s) = \nonumber \\ \frac{e}{mL} \dx_\nu(s)\int_s^\infty ds' e^{(s-s')/L}
\left(
F^{\mu\rho}_{ext}\dx_\rho\dx^\nu - 
F^{\nu\rho}_{ext}\dx_\rho\dx^\mu 
\right)(s')\nonumber \\
\ea
showing that the equation is cubic in the velocity (the position dependence will enter through the external electromagnetic field).

\section{Integral formulation of the $2^{\rm nd}$ Newton law} \label{sec:Newton}
While the Lorentz-Dirac equation is local in proper time, eliminating the initial condition with a future boundary condition turns-on an integration over future proper time in Eq.\eref{Rohrlich} and\eref{NewRadRecEq}.

This however is by itself not an obstacle. Newton's second law itself can be cast in the very same form by means of one integration, so that the velocity is related to an integral over all future times of the force. In this subsection we examine this formulation of this most basic  law of mechanics, in order to dispell some causality objections addressed here below and in section~\ref{subsec:causality} and to test numerical iterative methods later on in section~\ref{subsec:iterative}.

\subsection{Three-dimensional version}
Let us start this discussion by integrating the traditional form of Newton's second law in three-dimensions
\be \label{NewtonOld}
m\frac{d}{dt} {\bf v}(t) ={\bf F}(t)
\ee
in an analogous form to equation~(\ref{eqintermedia}),
\be\label{Newtonintermedia}
{\bf v}(t)= {\bf v}(t_i) +\frac{1}{m} \int_{t_i}^t dt' {\bf F}(t')\ .
\ee
At this point the velocity is expressed as an integral over the force at prior times, back to the time chosen for the initial condition, so causality is manifest. 

We now select those solutions that are asymptotically free, so that the velocity takes a constant value at large time, $\lim_{t\to \infty} {\bf v}(t)\equiv{\bf v_\infty}$.
Evaluating equation\eref{Newtonintermedia} at $t=\infty$ we see that
\be \label{eqforboundary}
{\bf v}(t_i)= {\bf v_\infty} -\frac{1}{m}\int_{t_i}^\infty dt' {\bf F}(t')
\ee
that expresses the velocity at time $t_i$ as an integral of the force over later times, in analogy  to 
equations\eref{Rohrlich} and \eref{NewRadRecEq} (except these provide the acceleration and not the velocity). 
This equation is obviously not in violation of causality, as one can reinterpret it easily as an initial value problem for $\bf v_\infty$.
Thus, one could understand equations such as\eref{Rohrlich} or \eref{NewRadRecEq} as integral initial value problems for an acceptable (not exponentially increasing) acceleration at infinite future time. 

The whole point of this exercise is that advanced effects do disappear by rewriting Newton's second law in integral form  in Eq.\eref{eqforboundary}
as an equation for ${\bf v_\infty}$.
If we had instead imposed as asymptotic condition that the particle be left at rest for $t=\infty$  ($\bf v_\infty=0$), we would have obtained
\be \label{Newton3D}
{\bf v}(t)=-\frac{1}{m}\int_{t}^\infty dt' {\bf F}(t')
\ee
an advanced equation such as Eq.\eref{Rohrlich} or\eref{NewRadRecEq} in apparent violation of causality (which it is not).

\subsection{Four-dimensional version of Newton's equation}\label{subsec:Newton4D}
Let us take now the relativistic generalization of Newton's second law in Eq.\eref{NewtonOld}
\be
m\ddx^\mu(s) = f^\mu (s)\ .
\ee
This can be multiplied by an integrating factor $\frac{1}{m} e^{-s/L}$ (here $L$ is any positive constant, not necessarily the electron relaxation length)  to yield
\ba
\frac{1}{m}e^{-s/L}f^\mu (s)& = &e^{-s/L}\ddx^\mu(s)  \\ \nonumber
&=& \frac{d}{ds}\left(e^{-s/L}\dx^\mu(s)  \right)+ \frac{1}{L} e^{-s/L} \dx^\mu(s)\ .
\ea
We can group two terms into an effective force per unit mass, 
$\frac{f^\mu (s)}{m}-\frac{\dx^\mu(s)}{L}$, and express the equation, integrating once with initial condition at $s_i$, as
\ba \label{intermediateNewtonRel}
e^{-s/L}\dx^\mu(s)= \\ \nonumber
e^{-s_i/L}\dx^\mu(s_i) +
\int_{s_i}^{s}ds'e^{-s'/L} \left( \frac{f^\mu (s')}{m}-\frac{\dx^\mu(s')}{L}
\right)
\ea
again in total analogy with Eq.\eref{eqintermedia}.
If we now restrict the solution space to those whose 
Lorentz dilatation factor $\gamma(v)=\left(\sqrt{1-v^2/c^2}\right)^{-1}$
cannot grow exponentially~\footnote{
This is a soft version of infinite-time asymptotic freedom, that in particular excludes the motion under a constant force extending to $t=\infty$.}, by imposing
$$
\lim_{s\to\infty}\left(e^{-s/L}\dx^\mu(s)\right) = 0 \ ,
$$ 
we can eliminate the initial condition by performing an evaluation of the integral equation\eref{intermediateNewtonRel} at infinity, and obtain
\be \label{NewtonRel}
\dx^\mu(s)= - \int_s^\infty ds'e^{(s-s')/L}
\left( \frac{f^\mu (s')}{m}-\frac{\dx^\mu(s')}{L}
\right)\ .
\ee
Once more we write it as an advanced integral equation that has several similarities with our Eq.\eref{Rohrlich}.
First, in both equations the highest order derivative of the position ($\dx$ in
Eq.\eref{NewtonRel}, $\ddx$ in Eq.\eref{Rohrlich}) appears on both sides of the equation. Second, the integrand in the right hand side of both equations has two terms, one depending on the external force and the other one on the movement of the particle
\footnote{The first objection could be lifted by integrating once Eq.\eref{NewtonRel} by parts to obtain
$$
\dx^\mu(s)= - \int_s^\infty ds'e^{(s-s')/L}
\left( \frac{f^\mu (s')}{m}-\frac{\x^\mu(s')}{L^2}
\right)\ .
$$
This is a bit closer to our equation Eq.\eref{NewRadRecEq} in that the highest derivative can be isolated on the left hand side, and it simplifies the numerical solution some.
}.

But Eq.\eref{NewtonRel} is different in one important count from Eq.\eref{Rohrlich}:
as we will show with several numerical examples, the velocity solving  Eq.\eref{NewtonRel} remains constant at any time for which the external force vanishes, while the acceleration solving Eq.\eref{Rohrlich} begins changing before the force (preacceleration).


\section{Further theoretical details}
\subsection{Problems with causality}\label{subsec:causality} 

The boundary condition in equation\eref{boundary2} and \eref{boundary3} guarantees that the solution
to the equation of motion does not grow up as $e^{s/L}$ at infinity, and this eliminates the self-accelerated solutions from the free Lorentz-Dirac equation (and from the integrodifferential equations derived thereof), when the external force vanishes for all times.

Nevertheless, what happens if the external force $f_{ext}$ does not always vanish? 

The issue of preacceleration has been often remarked.
Note from equations~(\ref{Rohrlich}) and~(\ref{NewRadRecEq}) 
that the acceleration at time $s$ depends on the force at later times $s'>s$.  Therefore, barring a built-in cancellation as happens in Newton's law in integral form, Eq.~(\ref{NewtonRel}), acceleration predates the force: 
 for a force vanishing before some given time, an observer that would have a time resolution of order $L$ could in principle see the particle accelerate before its action.  This will be visible in the numerical examples.

One often reads the argument that classical physics does not apply below the Compton scale, that is much bigger than 
the electron relaxation length
$$
\Lambda_{\rm Compton} = \frac{\hbar c}{m_e} \simeq 385 {\rm fm} \gg
L \simeq 1.876 {\rm fm} \ .
$$
While true, two observations are in order. First, from a more practical point of view, the scale $L$ in the equation of motion divides {\emph{proper time}} into units. From the point of view  of the laboratory, the relevant scale is the dilated distance $\gamma(v)L$.
For a 3 GeV electron typical for example of synchrotron light sources (such as ALBA~\cite{alba} in Barcelona, this factor is 5870, so that
$\gamma(v) L = 0.011$ nm. 
This means that the preacceleration can be a phenomenon at the atomic, not nuclear scale.

Second we should note that the electron relaxation length $L$ has dimensions of length and a fixed value independently of $\hbar$, while the Compton wavelength  depends linearly on $\hbar$.
It would be puzzling that if $\hbar$ happened to be a factor 1000 smaller, preaccelerations would manifest themselves at length scales larger than the new Compton wavelength. Quantum mechanical effects would not provide an escape in this case.


Let us also briefly discuss the Dirac-Lorentz equation in the following form
\be\label{DiracLorentzrewritten}
\ddx^\mu-L\dddx^\mu-\frac{f_{ext}}{m}=L\ddx^2 \dx^\mu\ .
\ee
If the radiation term in the right hand side was absent, this would be a linear inhomogeneous equation. Its general solution would have been an affine space spanned by the solutions of the homogeneous equation with $f_{ext}$ set to 0, added to any particular solution of the inhomogeneous equation. 
The self-accelerated solutions described in Eq.~(\ref{selfaccelerated}) are 
eliminated from the set of homogeneous solutions by setting the boundary condition at infinite time given by Eq.(\ref{boundary1}). This is sufficient to obtain only physical solutions.



When solving the equation of motion numerically, one might come across
an invariant interval $s$ with $f^\mu_{ext}=0$ but $\ddx^\mu\neq 0$. Then one knows that a self-accelerated mode is active. 
Since this is not a defect of our equation, but of the entire theoretical set up, we will continue with this warning in mind.

Some authors discussed in the past~\cite{Valentini,Goebel} whether the preacceleration phenomenon is related to  the lack of analyticity of the functions employed to expose the phenomenon (see for example fig.~\ref{fig:RadStep1} below).
The (shortened) argument is that the Lorentz-Dirac equation requires certain analyticity hypothesis not satisfied by such forces (so the claim is that they should not be used at all with the Lorentz-Dirac equation). With analytic forces the particle at times before the beginning of the force is not really preaccelerated by a future force, instead it reacts to the analytic extension of the future-time force to present times. We are not totally satisfied by this argument because Newton's equation in integrodifferential form, Eq.\eref{NewtonRel}, has a similar mathematical structure and does not present this behavior, as seen in Fig.~\ref{fig:Newtonstep}.

We have constructed another clarificatory example below in Eq.\eref{expinv}
that presents the unwanted preacceleration phenomenon while the force is exactly zero, although the function representing the force, given in Eq.~\eref{expinv} and shown in figure~\ref{fig:gaussian}, is continuous and all of its derivatives are continuous on the real line, even if it is not analytic in the complex plane. Still, it satisfies all conditions for the Lorentz-Dirac equation to be valid on the real line, thus weakening Valentini's argument 
about preacceleration being artificially induced by a poor mathematical choice for the external force.

\subsection{Further integrations}

The class of non-linear integral equations of motion that we consider here requires an iterative solution.
Once we have performed one iteration of the integral in the relevant equation of motion,  
for example  Eq.\eref{NewRadRecEq}, we have the highest derivative in hand for all times, for example
$\ddx(s)$.
This can in turn be immediately integrated to obtain the function and its lower derivatives, for which there is no inconvenience in setting initial conditions at initial time $\x(t_i)$, $\dx(t_i)$.

Thus, the velocity at arbitrary time is calculated from the acceleration as a simple quadrature
\be
\dx^\mu(s) = \dx^\mu(s_i) + \int_{s_i}^s ds' \ddx^\mu(s')
\ee
and the position is obtained in an analogous manner once the velocity is known,
\be
\x^\mu(s) = \x^\mu(s_i) + \int_{s_i}^s ds' \dx^\mu(s')\ .
\ee
Velocity and position will be updated iteratively at every step using these quadratures.

All functions would be known at this point in terms of the invariant interval $s$. To plot them in terms of laboratory time, all one needs to do is keep track, for each $s$, of the pairs $(\x^0(s),\ddx^\mu(s))$, $(\x^0(s),\dx^\mu(s))$, and $(\x^0(s),\x^j(s))$. The pair $(s,\x^0(s))$ itself provides the time-dependent Lorentz contraction factor $\gamma$ for all proper times and can be used to trade proper and laboratory time for each other.

\subsection{Small-$L$ approximation} \label{subsec:largemass}


The parameter $L$ is small, even for electrons, and more so for protons or heavier particles. Keeping only the zeroth and first orders in L, we can find interesting results to which we dedicate this subsection.

In this limit, Eq.\eref{NewRadRecEq} accepts a perturbative solution complementary to the iterative one that will be  pursued in subsection~\ref{subsec:iterative}.

Changing the variable under the integral sign in Eq.\eref{penultima} from $s'$ to $\sigma\equiv (s'-s)/L$, we have that the right hand side of the equation of motion $m\ddx^\mu = F^\mu(s)$ becomes
\be \label{auxHeavyM}
F^\mu(s) = \dx_ \nu(s) \int_0^\infty d\sigma e^{-\sigma} 
K^{\mu\nu}(s+L\sigma) 
\ee
with $K^{\mu\nu}$ defined in Eq.\eref{onlyK}.

If we expand around $L=0$, 
$$
K^{\mu\nu}(s+L\sigma) \simeq K^{\mu\nu}(s) + L\sigma \frac{dK^{\mu\nu}(s)}{ds}\dots
$$ 
that  we substitute into Eq.\eref{auxHeavyM}, keeping only the zeroth order and first orders, and employing $\int_0^\infty d\sigma e^{-\sigma}=1$,
this gives back
\be
m\ddx^\mu \simeq \dx_\nu K^{\mu\nu} + L \dx_\nu \dot{K}^{\mu\nu} 
\ee
which is a form of the Lorentz-Dirac equation where terms with both the third derivative and the square of the second derivative have been eliminated in terms of the external force.
Simple algebraic manipulations reduce it to the following form
\ba \label{nonrel}
m\ddx^\mu & = & f^\mu_{ext} + L\dx_\nu \left( 
\dot{f}_{ext}^\mu \dx^\nu -
\dot{f}^\nu_{ext} \dx^\mu \right) \\ \label{LandauLifschitz}
& = & f^\mu_{ext} + L  \left( 
\dot{f}_{ext}^\mu -(\dx \cd \dot{f}_{ext})\dx^\mu \right)
\ea
accurate to first order in $L$. This last equation seems to be a relativistic
version of the Landau-Lifschitz equation (see Eq.(36) of~\cite{Griffiths})~\footnote{We thank an anonymous referee for pointing this out to us.} valid for an arbitrary external force (the presentation of Landau and Lifschitz emphasizes electromagnetic forces only).

The first term returns the relativistic form of the second Newtonian law in the absence of radiation (e.g. neutral particle). 
Both terms are seen to be explicitly perpendicular to the four-velocity, as befits a relativistic force $dx^\mu f_\mu=0$. (this property is kept order by order in the $L$-expansion).
Eq.\eref{nonrel} is local; the acceleration at time $s$ depends on the external force and its time derivatives and the velocity of the particle at the same time $s$. There is no preacceleration (it vanishes also in the integrodifferential formulation when $L\to 0$).
As the second term of Eq.\eref{LandauLifschitz} is proportional to the external force (and its derivatives), there are no run-away solutions, unlike those found for the Lorentz-Dirac equation\eref{LorentzDirac}.

If higher order terms in $L$ are kept, one can express the radiation force on the particle in terms of its velocity,  the external force and its derivatives alone: the equation of motion remains quasilineal. 


In this same expansion, the radiated power can be written in terms of the external force alone as
\be
{\mathcal R}(s) = -\frac{L}{m} f^2_{ext}(s) - \frac{L^2}{m}\frac{df_{ext}^2(s)}{ds}\dots
\ee
(remember that the external force is a space-like four-vector, $f^2< 0$).

\section{Numerical method}\label{sec:nummethod} 

\subsection{Iterative approach} \label{subsec:iterative}

The iteration of Newton's second law in integral form\eref{NewtonRel} and also of the traditional equation of motion\eref{Rohrlich} presents a difficulty not affecting equation\eref{NewRadRecEq}.

To appreciate it, imagine applying the simplest approach to the iterative solution of the equation of motion (we will eschew Minkowski indices and all other unnecessary details, and refer simply to $\dx$ as the solution to the dynamical problem, as in Newton's case, but this might be $\ddx$ if attempting a solution of the more technically involved Dirac-Rohrlich equation\eref{Rohrlich}). 

In the naive-most approach, one simply guesses or calculates by other approximate means (e.g. the solution in the absence of forces, or the solution in the absence of radiation reaction, or an otherwise simple analytical case) a first guess
$\dx_{[0]}$. Then one plugs this guess into the right hand side of the integral equation to obtain 
$$
\dx_{[1]}(s) = \int ds' N(\dx_{[0]}(s'))
$$
where $N$  is the simplified form of 
\be
N^\mu(\dx(s')) \equiv e^{(s-s')/L}\left( \frac{\dx^\mu(s')}{L}-\frac{f^\mu(s')}{m} \right)
\ee
Iterating several times, the generic equation becomes
\be \label{update}
\dx_{[k]}(s) = \int ds' N(\dx_{[k-1]}(s')) \ .
\ee
If, to within some criterion (ours is specified below in subsection \ref{subsec:convergence}), convergence is reached, $\ar \ar \dx_{[k]}(s)-\dx_{[k-1]}(s)\ar \ar<\epsilon$, one would have achieved a solution of the integral equation by iteration.

However the appearance of the highest derivative on the right hand side of the older equation\eref{Rohrlich} brings about well-known instabilities, and this approach can fail. 
Consider the simpler force-free relativistic Newton's equation, that accepts a solution with arbitrary constant velocity (set by the initial condition). The iteration of
\be
\dx(s) = \int_s^\infty \frac{ds'}{L} e^{(s-s')/L} \dx(s')
\ee
with a constant solution becomes
$$
\dx_{[k]} = \dx_{[k-1]} \times \int_s^\infty \frac{ds'}{L} e^{(s-s')/L}\ .
$$
Analytically this equation is perfectly fine as the integral in the right hand side equals one. However, any finite-precision implementation on a computer means that the integral equals $(1-\epsilon)$ (with $\epsilon$ either positive or negative depending on the numerical integration algorithm).

Then one obtains
$$
\dx_{[k]} = \dx_{[k-1]} \times (1-\epsilon)
$$
that becomes, upon iteration, 
$$
\dx_{[k]} = \dx_{[0]} \times (1-\epsilon)^{k} \ .
$$
Depending on the sign of $\epsilon$, this converges (when the number of sweeps $k$ is sufficiently large) to  $0$ or diverges to $\infty$, with all other constant values being unstable points of the system. Thus, valid solutions are lost upon iteration.

Of course, this can be alleviated by Jacobi's method. To see it, let us recover the force and write the simpler relativistic Newton's equation\eref{NewtonRel} in discretized form as a linear system
\be
A \dx = b
\ee
with
\ba \nonumber
b_n = - \sum_{n'=1}^N e^{(s_n-s_{n'})/L} \frac{f_{n'}}{m} w_{n'} \\ \nonumber
A_{nn'}=\delta_{nn'}- \frac{w_{n'}}{L} e^{(s_n-s_{n'})/L} \ .
\ea
Here the integration weights have been denoted by $w_{n'}$.

Jacobi's method separates the diagonal piece $n=n'$, 
$$
D_{n} = \delta_{nn'}- \frac{w_{n}}{L}
$$
so that part of what would have been the right hand side in a plain iteration becomes part of the left hand side, yielding the vector equality
\be \label{Jacobi}
D\dx_{[k]} = b - (A-D) \dx_{[k-1]}
\ee
that is immediately solved by inverting $D$ (a diagonal matrix).
The convergence of this equation depends on the spectrum of $D^{-1}$.
If the integration weights $w_n$ are small (tightly spaced grid) then $\delta_{nn'}$ dominates in $D$, so its eigenvalues are near 1 and no quick divergence destabilizes the code. Cancellations between the terms on the right hand side of Eq.\eref{Jacobi}
are then facilitated and finite solutions can be found. Less favourable situations require more sophisticated algorithms.

These difficulties are a consequence of the highest derivative appearing inside the integral in the equation of motion. The modified equation\eref{NewRadRecEq} can be solved by simple iteration of the system and does not require the Jacobi method. However to stabilize the iteration we have found convenient to slow down the rythm of updates~\cite{Burden}, so that given $\dx_{[k-1]}$, and after calculating the right hand side, $\int K(\dx_{[k-1]})$, the update becomes, instead of Eq.\eref{update},
\be\label{update2}
\dx_{[k]} = a \int K(\dx_{[k-1]}) + (1-a) \dx_{[k-1]}\ ,
\ee
with the slow-down parameter $a$ typically 0.5-0.75.

\subsection{Computation of integrals over invariant interval}
The grid over the invariant interval is composed of equidistant points $s_1,s_2\dots s_N$. At the end of the computation the grid is translated (for plotting purposes) into laboratory time $t$, and it ceases being equispaced given that the time-dilation factor $\gamma(v)$ is variable.

To calculate the integral over the closed interval $(s,\infty)$, represented as $(s_j,\dots,s_N)$ on a discrete grid, we employ Simpson's rule (supplemented by a trapezoidal rule for the last integration interval if the number of points between $s_j$ and $s_N$ is even).

With finite computer time, $s=\infty$ must in practice be replaced by a cutoff $s_{max}$ on the maximum invariant interval of the particle. Since $\ddx(s\simeq s_{max})$ can no more be calculated from the integral equation (the later being advanced), the choice of the cutoff has to be such that this calculation is no more necessary. This can be achieved, for example, if the external force vanishes after a certain value of $s$. Then one expects that the particle will see vanishing accelerations at later times, and the acceleration ${\bf a}$ can just be set to zero in the vicinity of the cutoff without need of calculation. Each physical case requires separate examination here to set the boundary conditions.  This will be revisited in the next subsection~\ref{subsec:latetimes}

The number of points of the grid has to be such that $s_i-s_{i-1}< L \simeq 1.876$ fm, particularizing to an electron.
This guarantees that the fast-changing exponential factor $e^{(s-s')/L}$ is integrated over with sufficient accuracy.

 If one wants to follow the motion of the electron at large, superatomic scales at the nanometer or above, one then needs several million points on the grid for a slow electron. However,  radiation emission and radiation reaction are very small for slow electrons, and more interesting is the case of electrons with $\gamma$ factors of order 5-10 (emitted by beta decay and other nuclear processes, or accelerated by small machines) up to the thousands (for synchrotron light sources) or hundreds of thousands (for colliders at the energy frontier). Since $s_{max}=t_{max}/\bar{\gamma}$ for some average Lorentz contraction factor on the trajectory, to reach the nanometer scale in the laboratory the grid in $s$ can be much coarser. We use steps of order 0.1-1 fm in $s$ and our calculations typically require 10 000 points as an order of magnitude estimate. 
Such calculations can run on a typical 2-3 GHz processor in few minutes. Large computations running on a dedicated cluster or supercomputer can potentially reach human size scales even with fermi steps in invariant interval.

\subsection{Behavior at large times, 
initial condition}\label{subsec:latetimes}

Since the integral equation is built with a boundary condition at infinite time, it is worth delving on the behavior of the computer code at large times.

Vlasov~\cite{Vlasov:1997rt} has observed that the solution with preacceleration present and $\ddx^\mu$ finite as $s\to\infty$ is unstable to perturbations of the acceleration in the future. Our experience with the system confirms that setting and maintaining the boundary condition at infinity requires some care in the numerical analysis.

Checking first Newton's Eq.\eref{NewtonRel}, it is easy to note that if we set $f_{ext}=0$ at large times (asymptotically free-particle condition), then the zeroth component of the velocity satisfies
\be
\dx^0(s) = \int_s^\infty ds' \frac{\dx^0(s')}{L} e^{(s-s')/L}
\ee
(for large $s$) and thus a constant solution is possible, and after pulling $\dx^0$ out of the integral, shifting this, and using 
$$
\int_0^\infty \frac{d\tau}{L} e^{-\tau/L} = {\rm constant} =1\ , 
$$
we obtain that $\dx^0=\gamma$ (constant) at large time, so that the coordinate time scales linearly with invariant interval as expected.  The same reasoning holds for the space-like components of Eq.\eref{NewtonRel}, yielding a constant velocity at infinite time, as expected from Newton's first law. 

To implement this behavior on the computer, all we do is require, after reaching a certain $s_{free}$ large enough, to read-off the last calculated point and prolong the solution continuously to the cutoff $s_{max}$, so that
\be
\dx\left( s\in(s_{free},s_{max})\right) = \dx(s_{free})\ .
\ee
In the next iteration, we start computing the integral right hand side of Eq.\eref{NewtonRel} for $s\le s_{free}$, and at the end, set again the last points in the large-$s$ interval equal to the new value at $s_{free}$.  

The initial condition drifts from iteration to iteration in Jacobi's method
and needs to be reimposed in the iterative process.
We do this by rescaling the entire function, be it by employing a linear shift or a dilatation, depending on whether the initial condition was or was not zero. The rescaling stops being necessary once convergence has been achieved, so that the integral equation and the initial condition have simultaneously been satisfied.


The advantages of the modified equation (that accepts a simple iterative solution such as Eq.\eref{update2}) come again to the forefront; all we need to do to impose initial values of position and velocity (and zero acceleration in the absence of forces), is to set $\x_{[0]}(s_1)$
$\dx_{[0]}(s_1)$, and $\ddx_{[0]}(s_1)$ in the first iteration, and not recalculate them. The program keeps their value for its entire duration.

\subsection{Convergence criterion}\label{subsec:convergence}

At each iteration $[n]$ we need to decide whether the solution is converging (the integral equation at hand is closer to being satisfied) or not. 
As a diagnostic number we employ an integral measure for each of the position and velocity (for Newton's Eq.\eref{NewtonRel} in relativistic form) and an additional one for the acceleration (for Eq.\eref{NewRadRecEq}).

For example the convergence function for the velocity is, assuming $s_i=0$, 
\be
{\mathcal C}_{\dx} = \frac{\left( \int_0^\infty  ds' \sum_{\mu=0}^3  
\left( \dx_{[n+1]}^\mu(s')- \dx_{[n]}^\mu(s')\right)^2 \right)^{1/2} }{\left( \int_0^\infty  ds' \sum_{\mu=0}^3  
\left( \dx_{[n+1]}^\mu(s')+ \dx_{[n]}^\mu(s')\right)^2 \right)^{1/2} }
\ee
(where we are \emph{not} employing the Minkowski metric, but a sum over absolutely positive quantities). 
In the limit $n\to\infty$, one would expect, if the numeric method approaches convergence, that ${\mathcal C}_{\dx}\to 0$. In practice this is reached to very good accuracy after several dozens of iterations. 

We find that the iteration does not tend to drift between solutions (a common problem in numerical analysis), but it may develop instabilities and lose all contact with an approximate solution of the integral equation. In such case, one must stop execution of the program in an orderly manner by testing whether ${\mathcal C}_{\dx}$ is near 1 (since the criterion is constructed as a ratio of difference over sum, this means that the two functions $\dx_{[n+1]}^\mu$ and
$\dx_{[n]}^\mu$ are not commensurate and one is dominating the difference in the numerator, which is a signal of numeric divergence). 
After the stop, the program is ran again with a different initial guess or a tighter grid.

\subsection{Parameter relaxation}

Since the equation of motion is non-linear in $\dx$, often a solution will not be easily found for a given force. This is for example the case for oscillatory forces that have complicated solution functions, such as the particle in a magnetic field that we study below.

A workable strategy to find such solutions is to select a physical or a grid parameter and start with a trivial value. For example, extend the grid only as far as the first oscillation of the force, and for larger $s$ cutoff the force. One can then increase the cutoff in small steps to include more and more of the force.

Another example is the parameter $L$, in certain cases it is convenient to work with very large or very small $L$, and then relax it to its physical value. 

In either case, one lets the computer code find a fully converged solution for a given $L$ or $s_{max}$, then varies this parameter slightly, and find convergence again, procedure iterated as many times as needed until the necessary value of the parameter is reached. We have used this method extensively in the computations in section~\ref{sec:Numeric}.

\section{Numerical examples} \label{sec:Numeric}

In this section we present several numerical examples. 
We will use a system of units with $\hbar=c=1$, and length measured in fm. Also 
${\bf F}$ will represent the force per unit mass in the laboratory frame.

Subsection~\ref{subsec:solveNewton} serves to gain experience with the algorithm with Newton's equation in integral form, be it the standard three-dimensional or the covariant form in Minkowski space. 

In subsection~\ref{subsec:solveElementary} we will consider several examples with radiation reaction: in each one, we define a force ${\bf F}$ in the laboratory frame, iterate the covariant equation\eref{NewRadRecEq} until the desired convergence is reached~\footnote{Note that $f^i=\gamma F^i$, $f^0=\gamma {\bf F}\cd{\bf v}$, and therefore at each step of the iteration the external force depends on the velocity 
$\dx^\mu$ obtained in the previous step, until convergence is attained.} and then we represent the results in the laboratory frame with the help of relations\eref{acceleration} and\eref{acceleration2}. For simplicity we will consider first a step function, and then an oscillatory force.
In all cases represented in this section, the force plotted is the force per unit mass.

Subsection~\ref{subsec:solvelarger} demonstrates that the equation and algorithm can be scaled to larger systems much beyond the few-Fermi scale, showing their practical use for real problems.

\subsection{Solution to Newton's equation in integro-differential form}
\label{subsec:solveNewton}
\begin{figure}
\includegraphics*[width=8.5cm]{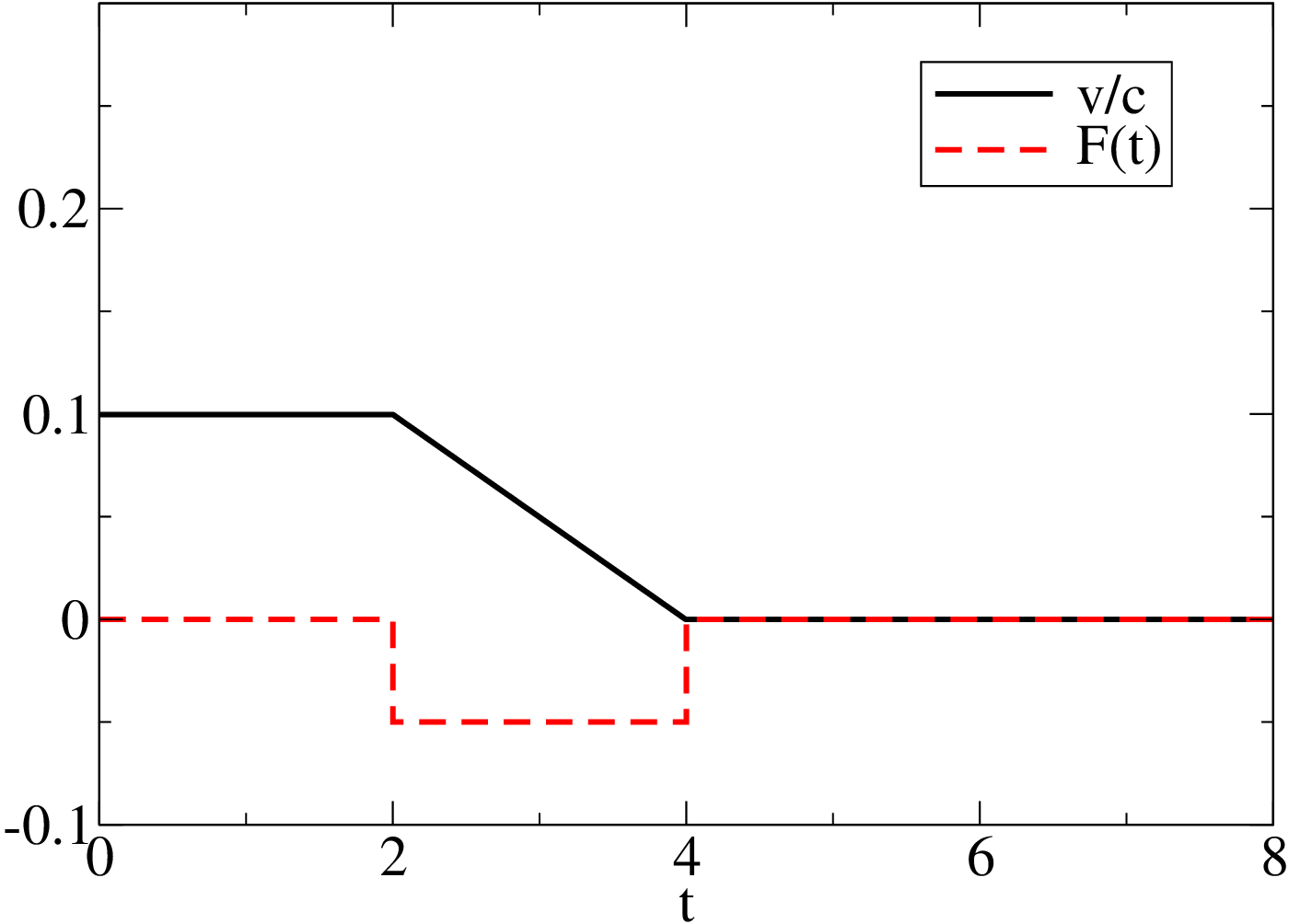}
\caption{Solution of Newton's equation in integro-differential form in three dimensions, Eq.~(\ref{Newton3D}) for a step force (dashed line). We plot the velocity (solid line). The numerical solution coincides with the expected one imposing the velocity to vanish at infinity as boundary condition. (Arbitrary units of length).}
\label{fig:Newton3Dstep}
\end{figure}

As a test of our numerical iteration algorithm, let us take
as a first example the three-dimensional second law of Newton in equation\eref{Newton3D} for a square well repulsive potential, depicted in figure \ref{fig:Newton3Dstep}.
The force (in the z-direction) can be written down in terms of two step functions as
\be
F_z = -f \left( \theta(t-t_0)-\theta(t-t_1)\right)
\ee
and the general analytical solution is
\ba
v_z(t) = v_0 - \nonumber \\ \nonumber f \left[
(t-t_0)  \left(\theta(t-t_0)-\theta(t-t_1)\right)
+(t_1-t_0)\theta(t-t_1) \right] \\
\ea
(remember that we use the force per unit mass).

This solution for the velocity (incidently, showing no preacceleration) for the particular values $f=1$ and (as built-in in Eq.\eref{Newton3D}) imposing
$v_\infty=0$, is computed iteratively with Eq.\eref{Newton3D} and plotted in the figure, coinciding with the exact analytical solution~\footnote{
Since we have an explicit exact solution, it will satisfy the integral equation identically, except for round-off errors. If we begin with an initial ansatz that is not close to the actual solution, convergence is attained in a few tens of iterations.
}.

\begin{figure}
\includegraphics*[width=8.5cm]{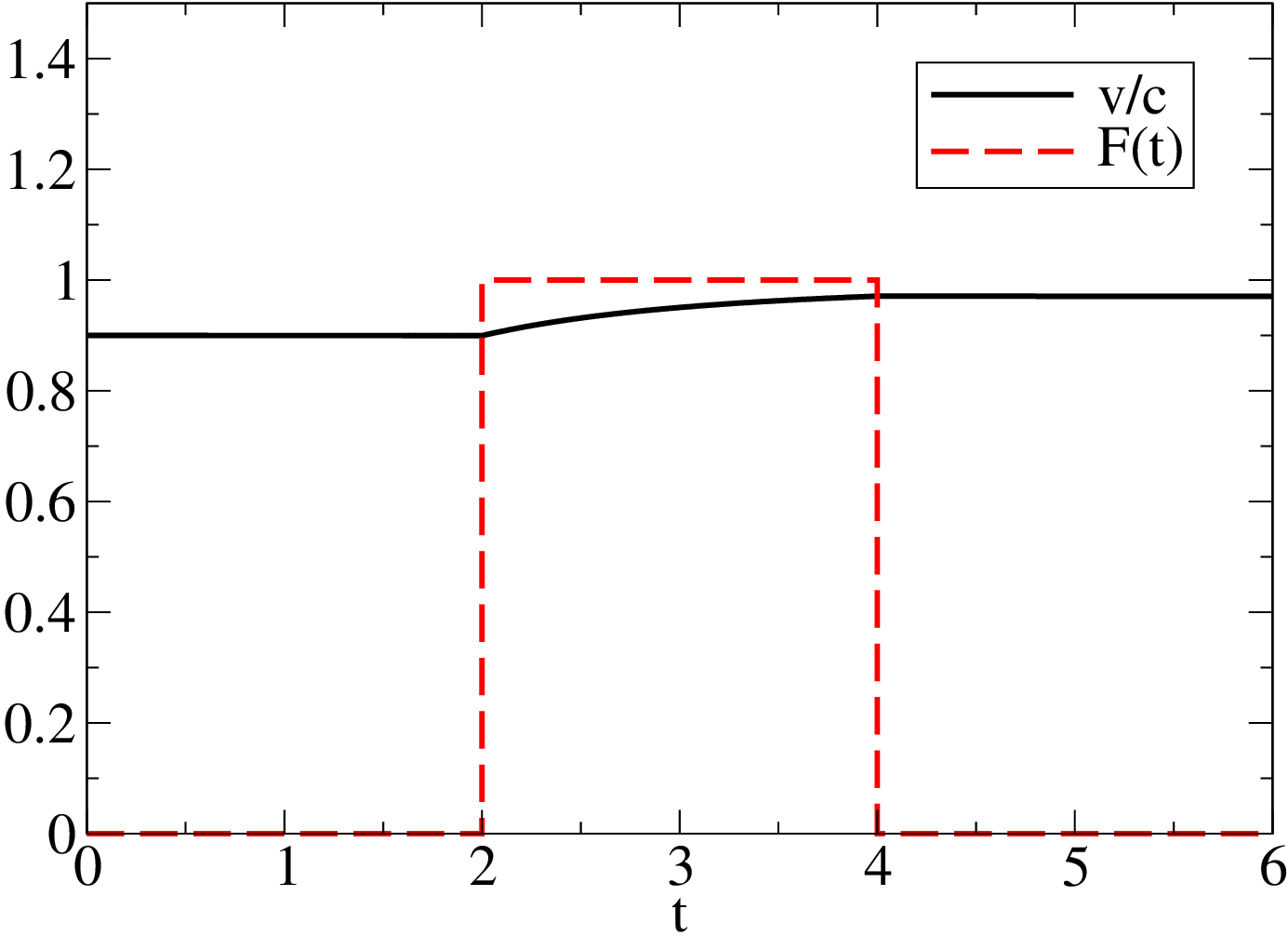}
\caption{Solution of the relativistic version of Newton's equation in integro-differential form in four dimensions, Eq.~(\ref{NewtonRel}) for a step force (dashed line). We plot the velocity (solid line). The numerical solution coincides again with the expected one, and in particular there is no preacceleration.(Arbitrary units of length).}
\label{fig:Newtonstep}
\end{figure}

Next we perform the same computation with the relativistic Newton law, Eq.\eref{NewtonRel}, and plot the result in figure~\ref{fig:Newtonstep}. Again causality is manifest and there is no preacceleration. We set as initial condition ${\bf v}(t_i)=0.9c{\bf e}_{z}$ and the force parallel to the velocity.

\begin{figure}
\includegraphics*[width=8.5cm]{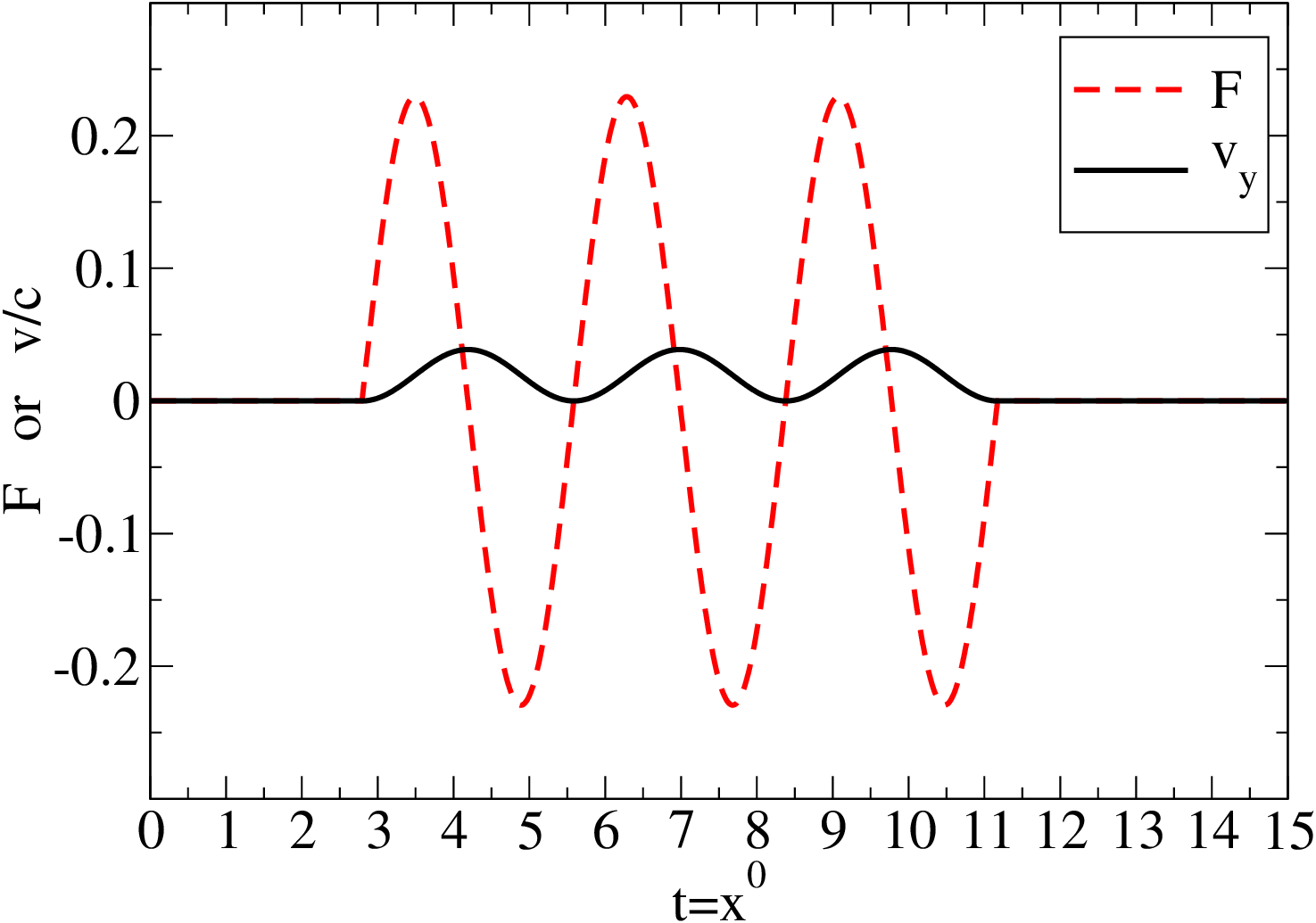}
\caption{Solution of the relativistic version of Newton's equation in integro-differential form in Minkowski space, Eq.~(\ref{NewtonRel}) for an oscillating force (dashed line). We plot the velocity (solid line).  There is no preacceleration. In fact, Newton's first law is fully satisfied, the velocity being extremum at the zeroes of the force. (Arbitrary units of length).}
\label{fig:Newtonoscillatory}
\end{figure}

Finally we solve again Eq.~(\ref{NewtonRel}) with a periodic potential truncated at two  times $t_0$ and $t_1$, with the result visible in figure~\ref{fig:Newtonoscillatory}. 
The initial velocity is again parallel to ${\bf e}_{z}$ and the oscillatory force is now perpendicular to this, along the $OY$ axis.\\
Again there is no preacceleration. Moreover, Newton's first law is manifest, as it is seen that every time for which the force vanishes features also an extremum of the velocity (vanishing acceleration). 

In conclusion of these first exercises, there is nothing unexpected in casting and solving Newton's law in either three-dimensional or Minkowski form as an integrodifferential equation with an advanced boundary condition at $t=\infty$.

\subsection{Elementary solutions with radiation reaction}
\label{subsec:solveElementary}

Studies of radiation reaction have a long history. An interesting early paper with numerical and analytical solutions compiled together is due to Plass~\cite{Plass:1961zz}.

In line with this classic work we now proceed to show example solutions of Eq.\eref{NewRadRecEq}. The components of the velocity and acceleration in the laboratory shown in the figures have been obtained from the four-dimensional covariant quantities that appear in that equation through the relations
\ba \label{acceleration}
v^i &=& \frac{\dx^i}{\dx^0} \\ \nonumber \\ \label{acceleration2}
a^i &=& \frac{\ddx^i-\ddx^0  v^i}{\gamma^2} \ .
\ea

%

\begin{figure}
\includegraphics*[width=8.5cm]{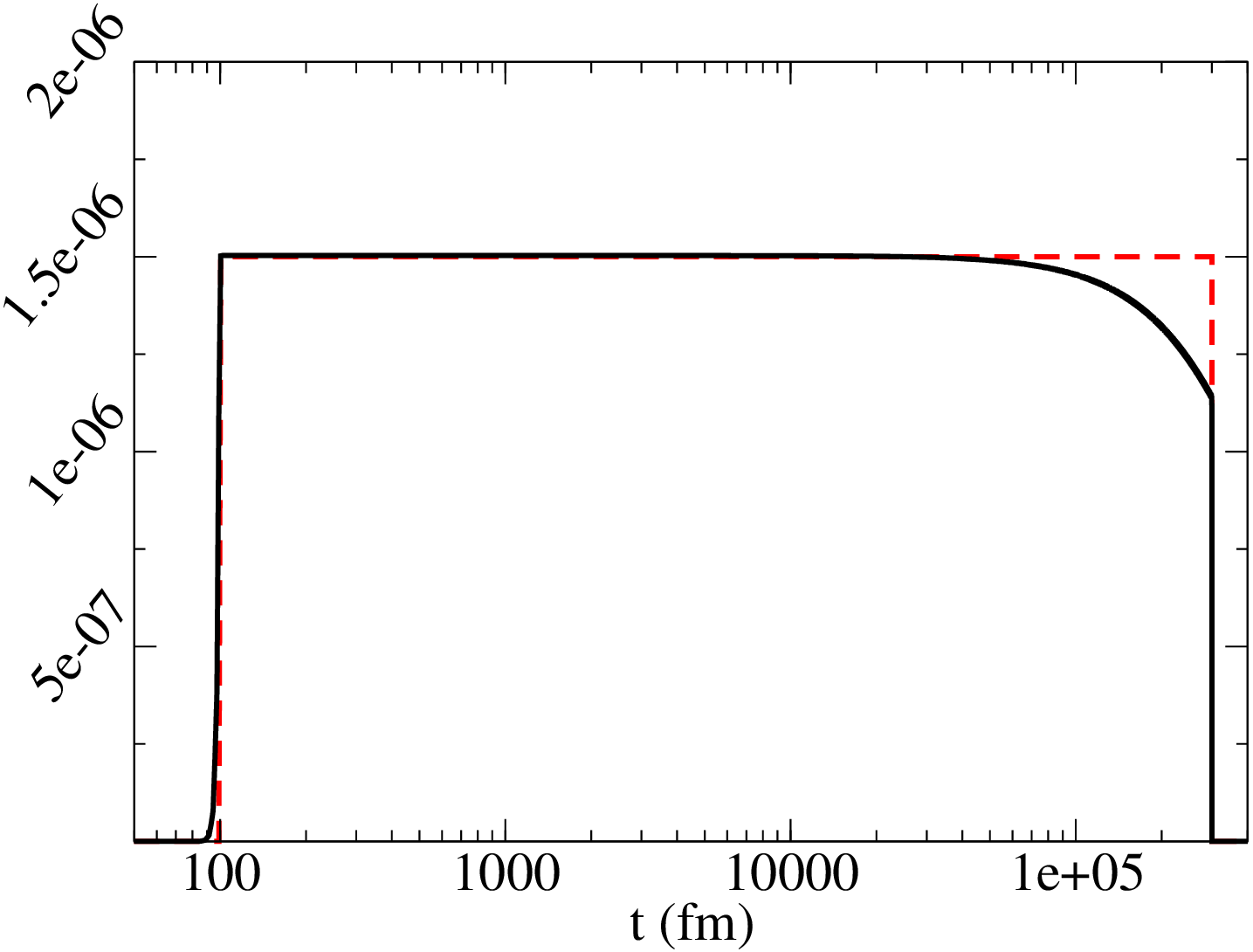}
\includegraphics*[width=8.5cm]{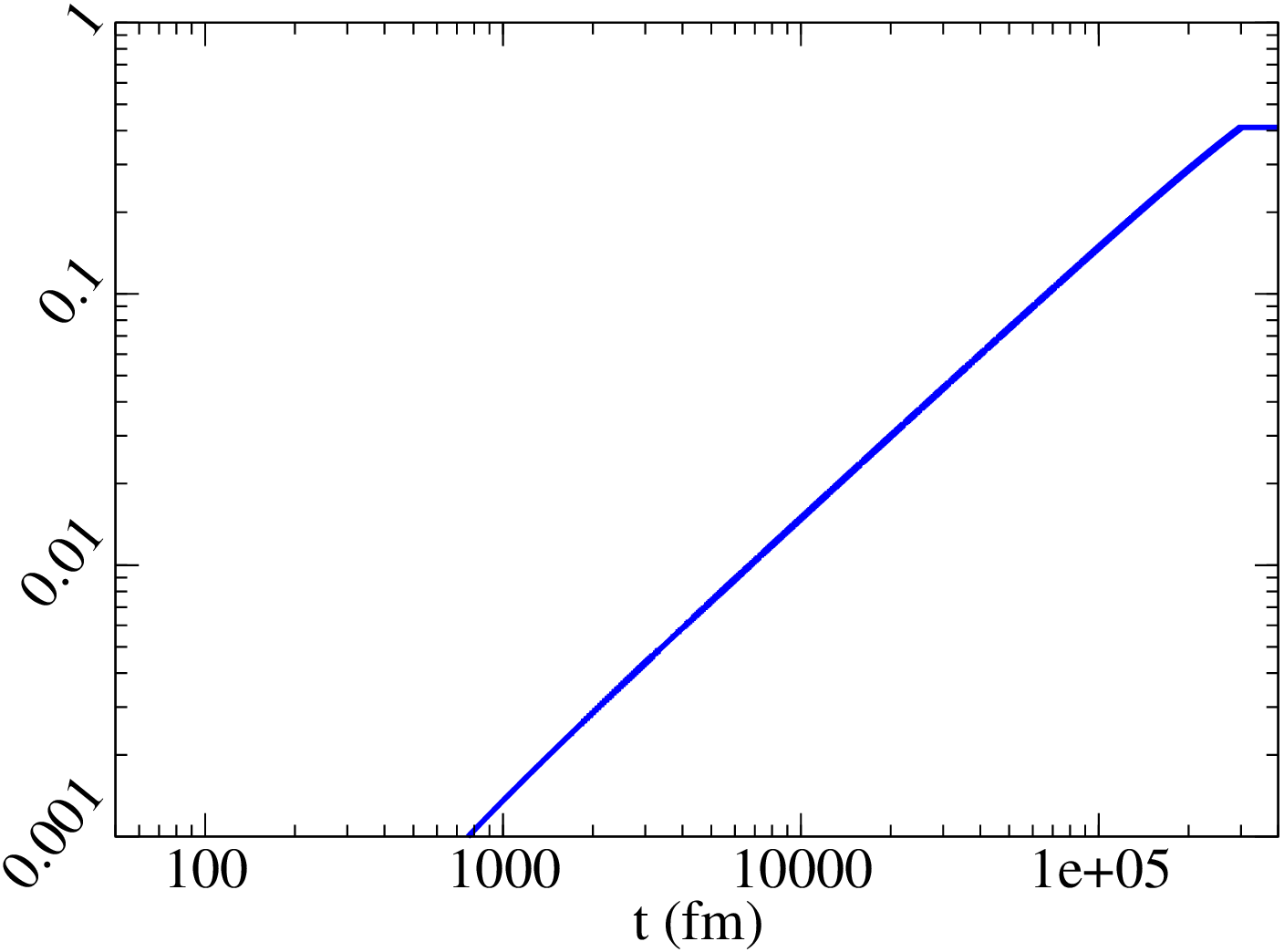}
\caption{
Solution to the modified equation of motion with radiation reaction, Eq.~(\ref{NewRadRecEq}) for a feeble step force (top plot, dashed line: modulus of the three-force in the laboratory; solid line: acceleration).  
While the velocity is far from $c$ (bottom plot), the acceleration is practically constant. It diminishes slowly as the particle becomes relativistic.
 We see distinctly preaceleration. The positive acceleration decreases rapidly before the force stops,
but due to the logaritmic scale this effect is not visible in the figure; we will show it in linear scale in fig.~\ref{fig:RadStepFlatClose}.
(Units of length are fm; note that the $OY$ axis represents inverse fm for the acceleration and for the force per unit mass.)}
\label{fig:RadStepFlat}
\end{figure}
%

The calculation is reported in 
figure~\ref{fig:RadStepFlat}. The iteratively calculated solution of Eq.\eref{NewRadRecEq} shows the following. The particle begins accelerating before the force starts.
%
Once it does start acting, the acceleration is almost constant and approximately equal to the external force per unit mass, the energy radiated being very small
\footnote{It is well known that the Lorentz-Dirac differential equation has analytical solutions with constant acceleration under a constant force in the proper reference system (hyperbolic motion), for which the last two terms in Eq.\eref{LorentzDirac} cancel mutually in an exact way. Here, in the iterative solution of our Eq.\eref{NewRadRecEq}, this constancy of the acceleration in the Laboratory frame at the beginning of the action of the force is only approximate, and it stays so only as long as the dilatation factor $\gamma$ remains near one, and the motion under a constant force in the Laboratory almost coincides with an exact hyperbolic motion in the proper system}. 
\begin{figure}
\includegraphics*[width=8.5cm]{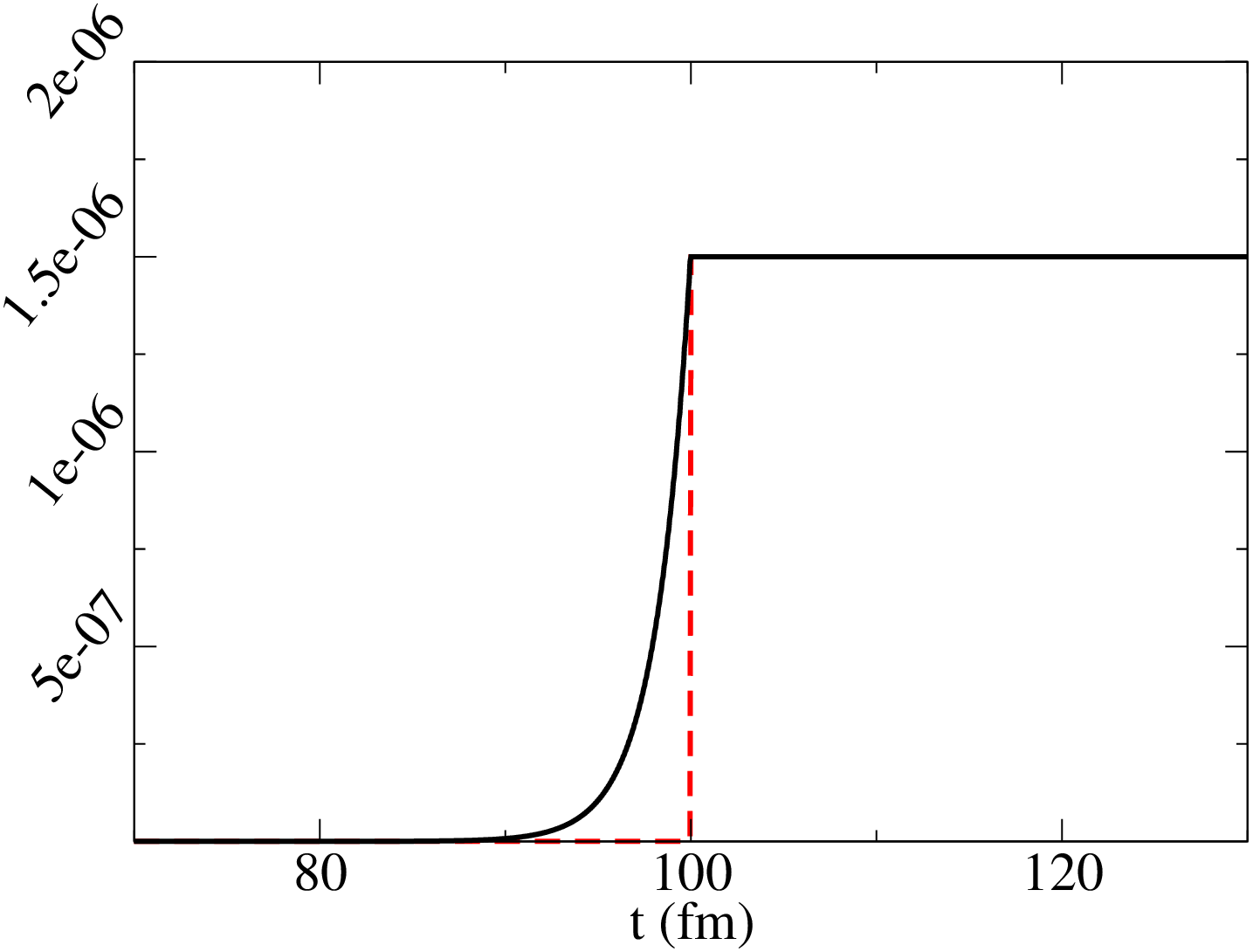}
\includegraphics*[width=8.5cm]{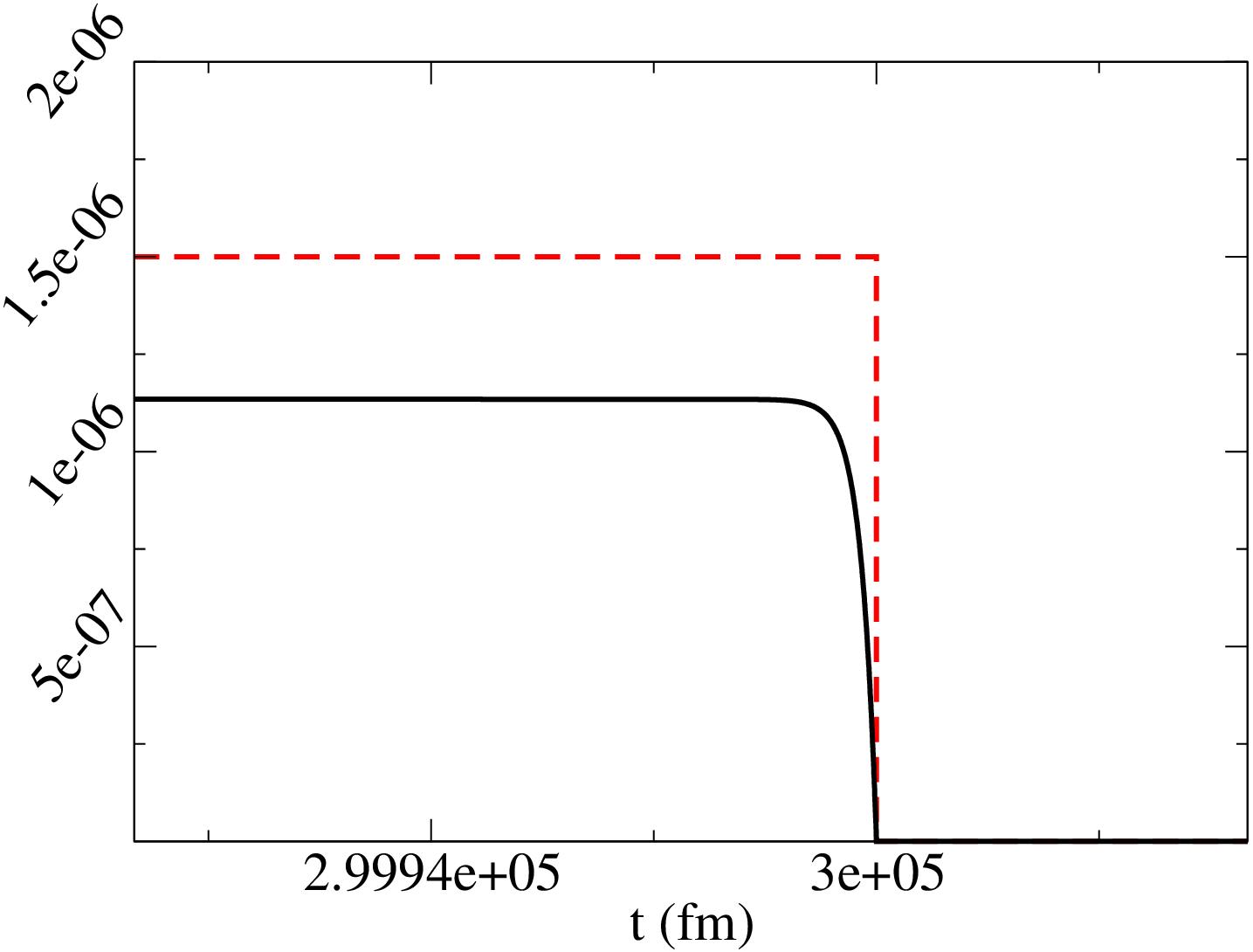}
\caption{
Detail of figure \ref{fig:RadStepFlat} where we show smaller time intervals around $t_0$ and $t_1$, to discern the effects of preacceleration.
These effects are a consequence of the advanced integration in Eq.\eref{NewRadRecEq}.
}
\label{fig:RadStepFlatClose}
\end{figure}

Further on, if the velocity in the Laboratory tends to one, the acceleration must necessarily go to zero; but if the external force stops before approaching this asymptotic limit, as in figure \ref{fig:RadStepFlat}  the acceleration falls suddenly to zero before the force ends, a phenomenon visible in some of the figures of Plass~\cite{Plass:1961zz} (although traditionally emphasis was put only on the preacceleration). Finally, after the external force ends the acceleration vanishes, as it is evident looking to Eq.\eref{NewRadRecEq}.

In figure \ref{fig:RadStepFlatClose} we show the preacceleration effect with more detail by looking closely to the times $t_0$ and $t_1$ where the force starts and ends its action.

If classical electromagnetic theory could be applied to forces very intense, acting during very short times, the oddities would manifest themselves in a more extreme way: we show them in the following examples as an extreme manifestation of the  mathematical properties of the formalism, \emph{without any pretension of physical application}.

\begin{figure}
\includegraphics*[width=8.5cm]{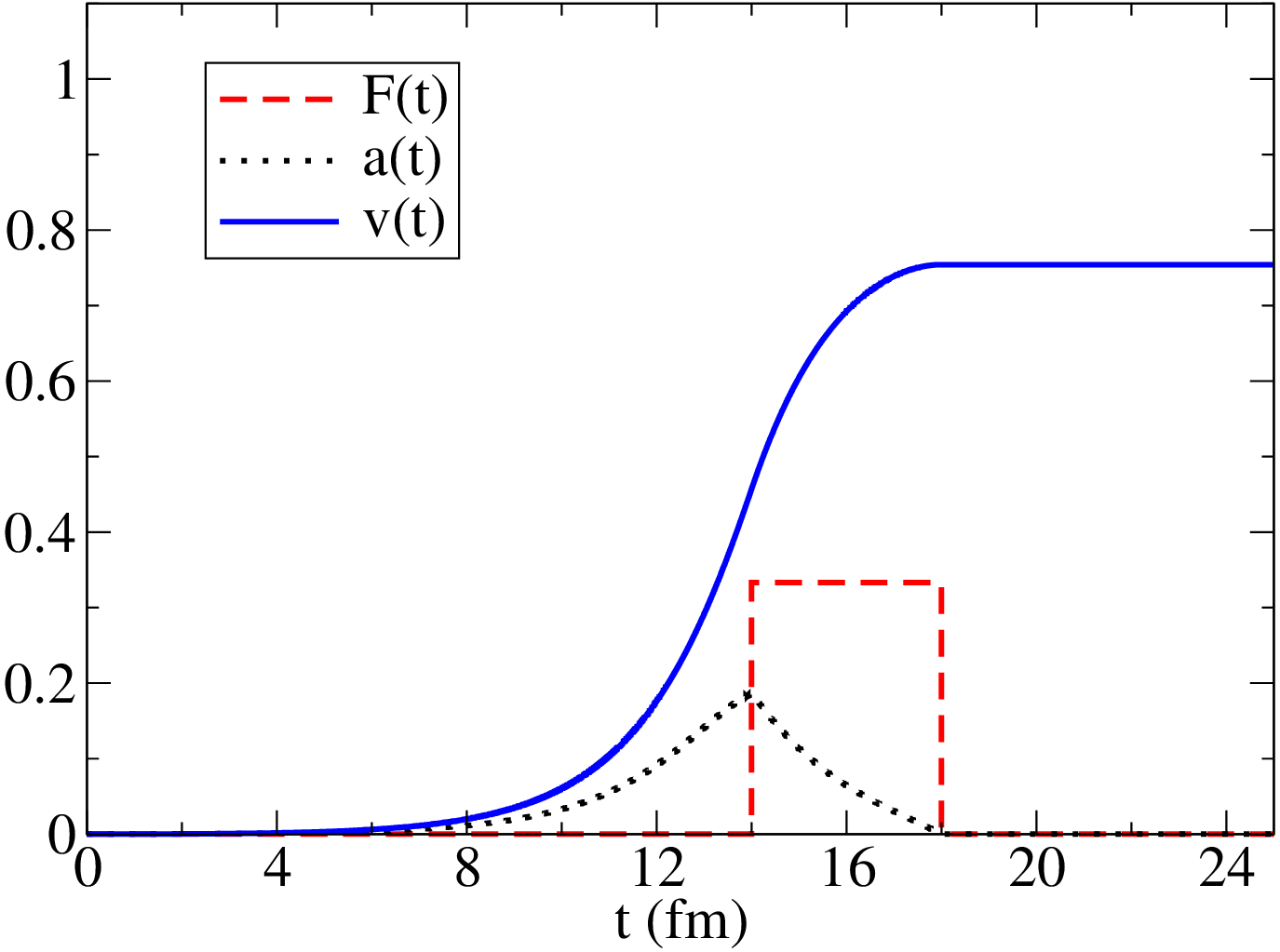}
\includegraphics*[width=8.5cm]{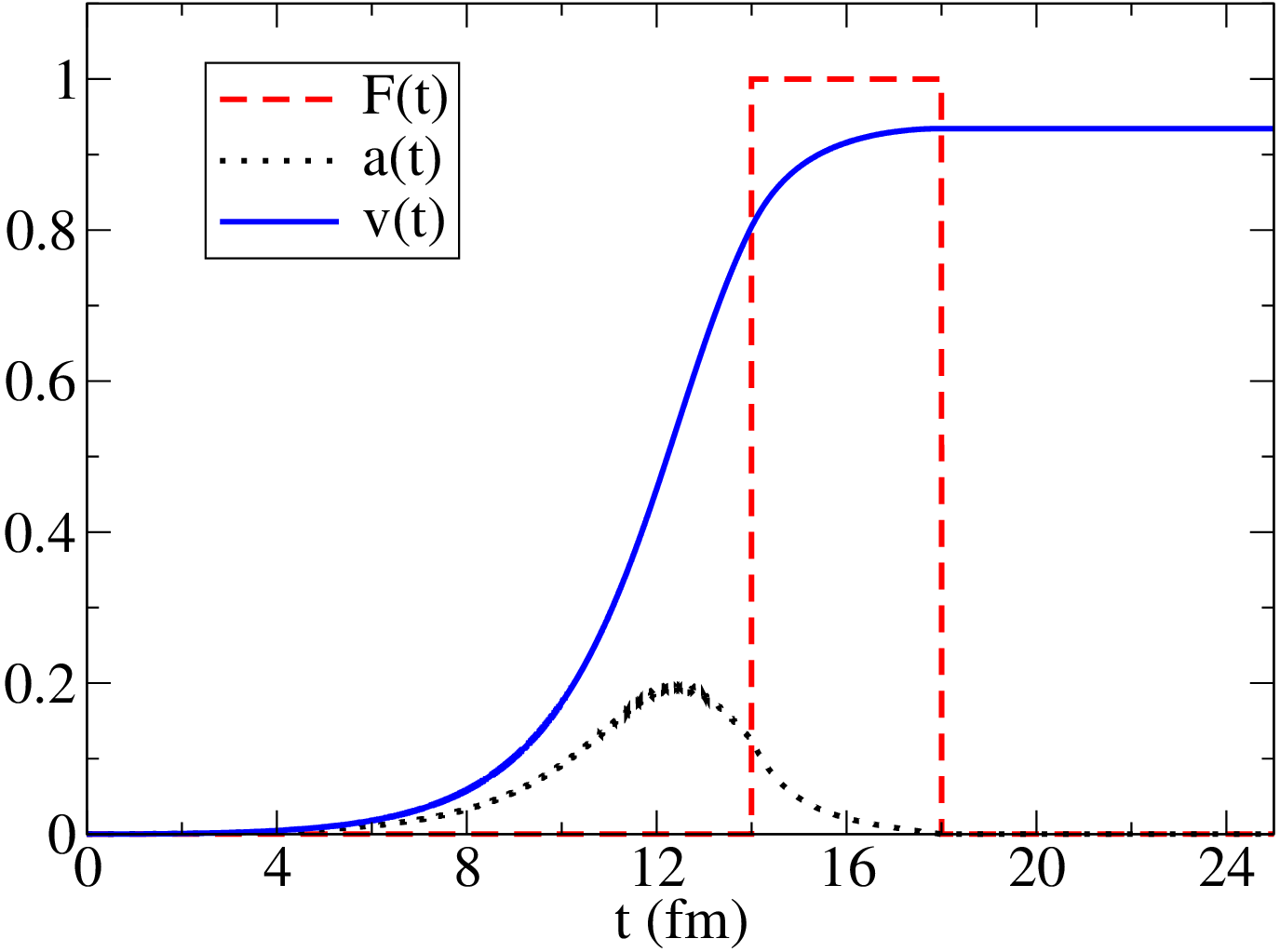}
\caption{Solution to the modified equation of motion with radiation reaction, Eq.~(\ref{NewRadRecEq}) for a step force (dashed line: modulus of the three-force). The acceleration (solid line)  takes off at a time of order $L=2e^2/(3m_ec^2)$ before the actual application of the force, in violation of causality, as is well known from other formulations of the theory. 
The particle is initially at rest. The force in the bottom plot triples the intensity of the force in the upper plot. 
 (Units as in figure~\ref{fig:RadStepFlat}.)}
\label{fig:RadStep1}
\end{figure}

\begin{figure}
\includegraphics*[width=8.5cm]{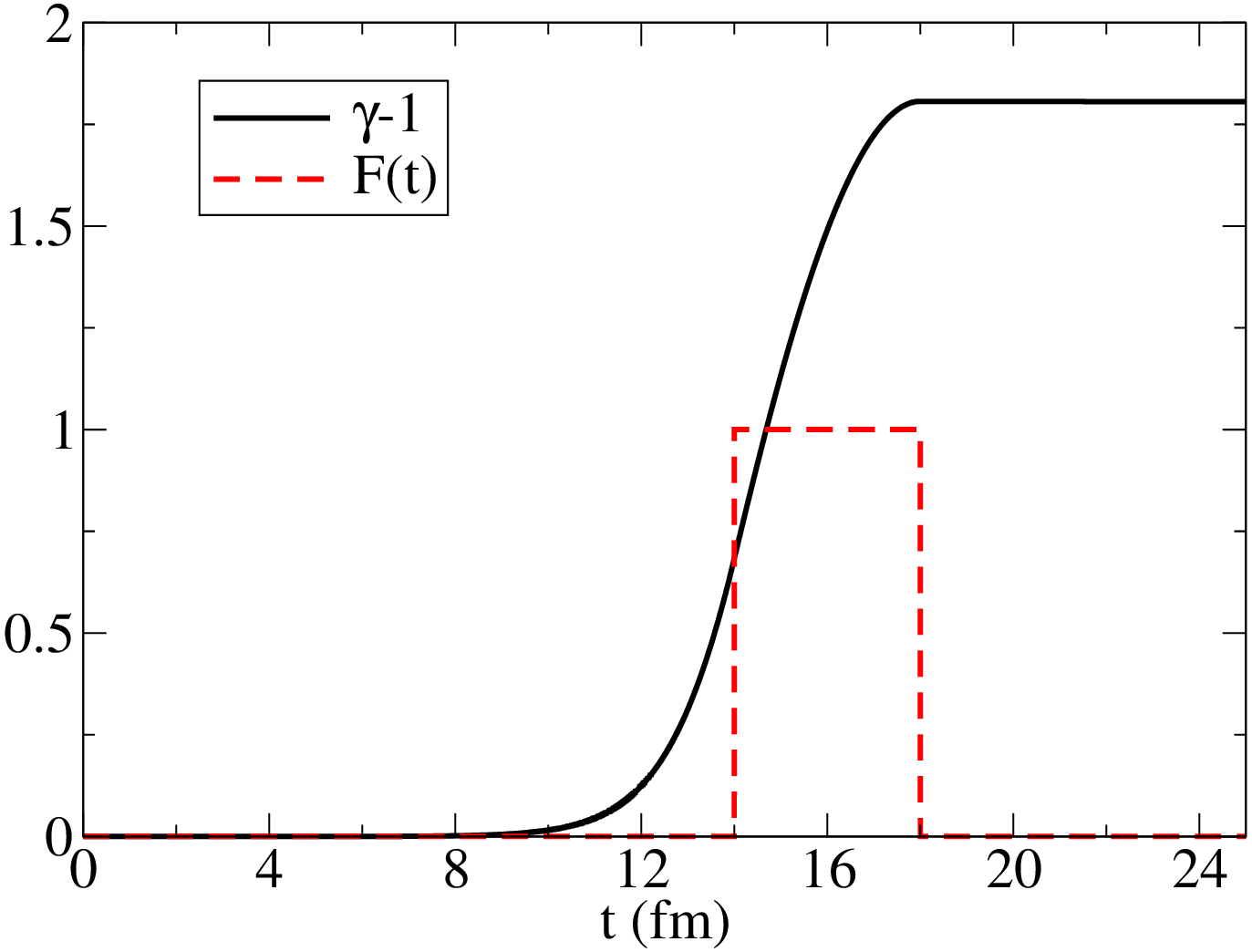}
\caption{
Kinetic energy per unit mass $\gamma(v(t))-1$ (solid line) for a square-hat force that acts on a finite interval of time, same as bottom panel of figure~\ref{fig:RadStep1}.}
\label{fig:RadStep1bis}
\end{figure}

In figure~\ref{fig:RadStep1} we employ again a step-force, affecting a particle initially at rest. The advanced features are obvious in that 
preacceleration is present before the force acts,
but also the acceleration is dropping quite fast well before the action of the force stops.
In the bottom panel, it even starts dropping {\emph{before}} the force acts. 
This last effect is a consequence of relativity, as the particle's velocity in the laboratory approaches $c=1$, the laboratory acceleration needs to vanish.
To show this we plot in figure~\ref{fig:RadStep1bis} the relativistic kinetic energy per unit mass, $T/m = \gamma(v(t))-1$, where we see that the energy increases steadily.

\begin{figure}
\includegraphics*[width=8.5cm]{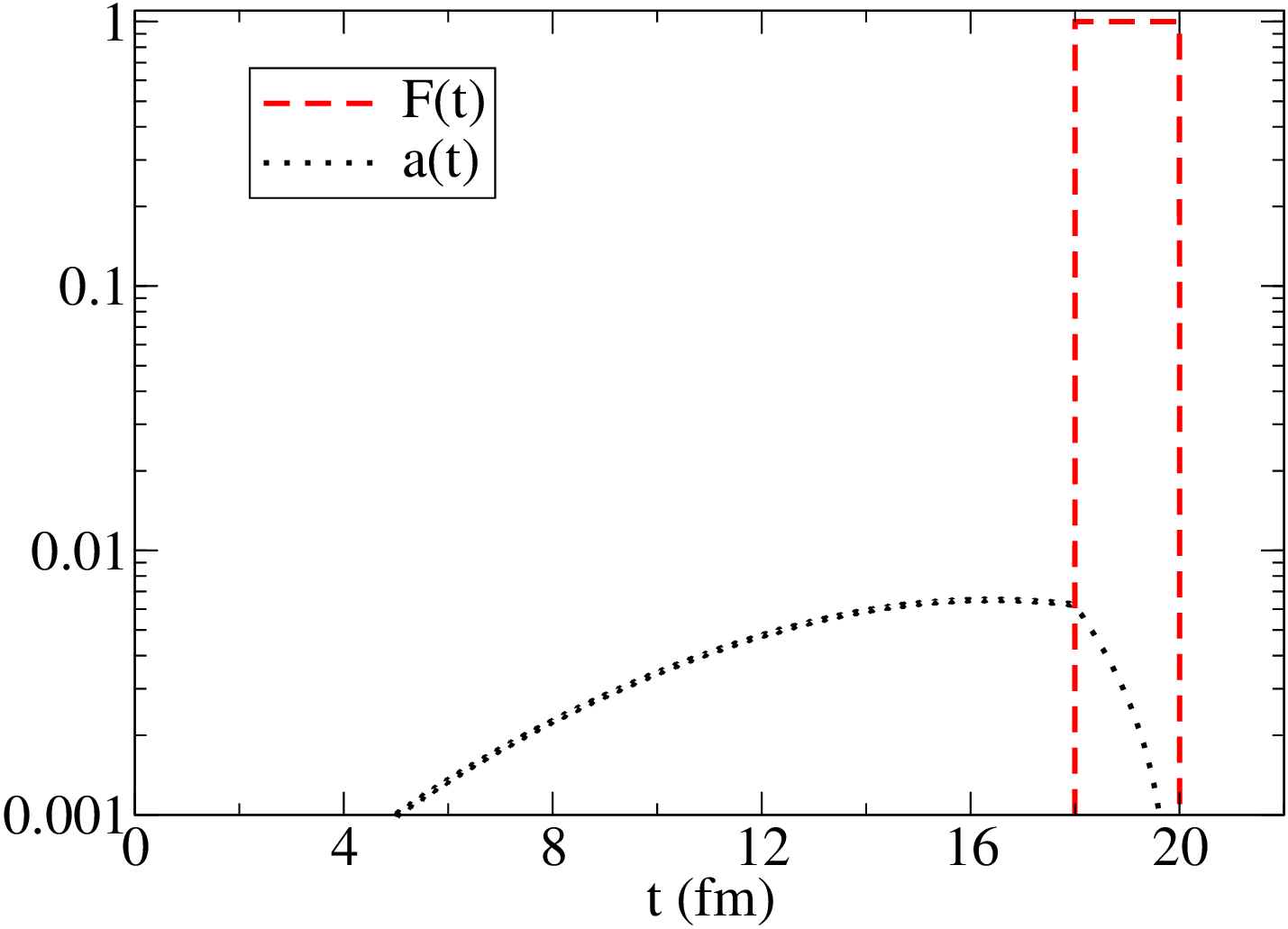}
\includegraphics*[width=8.5cm]{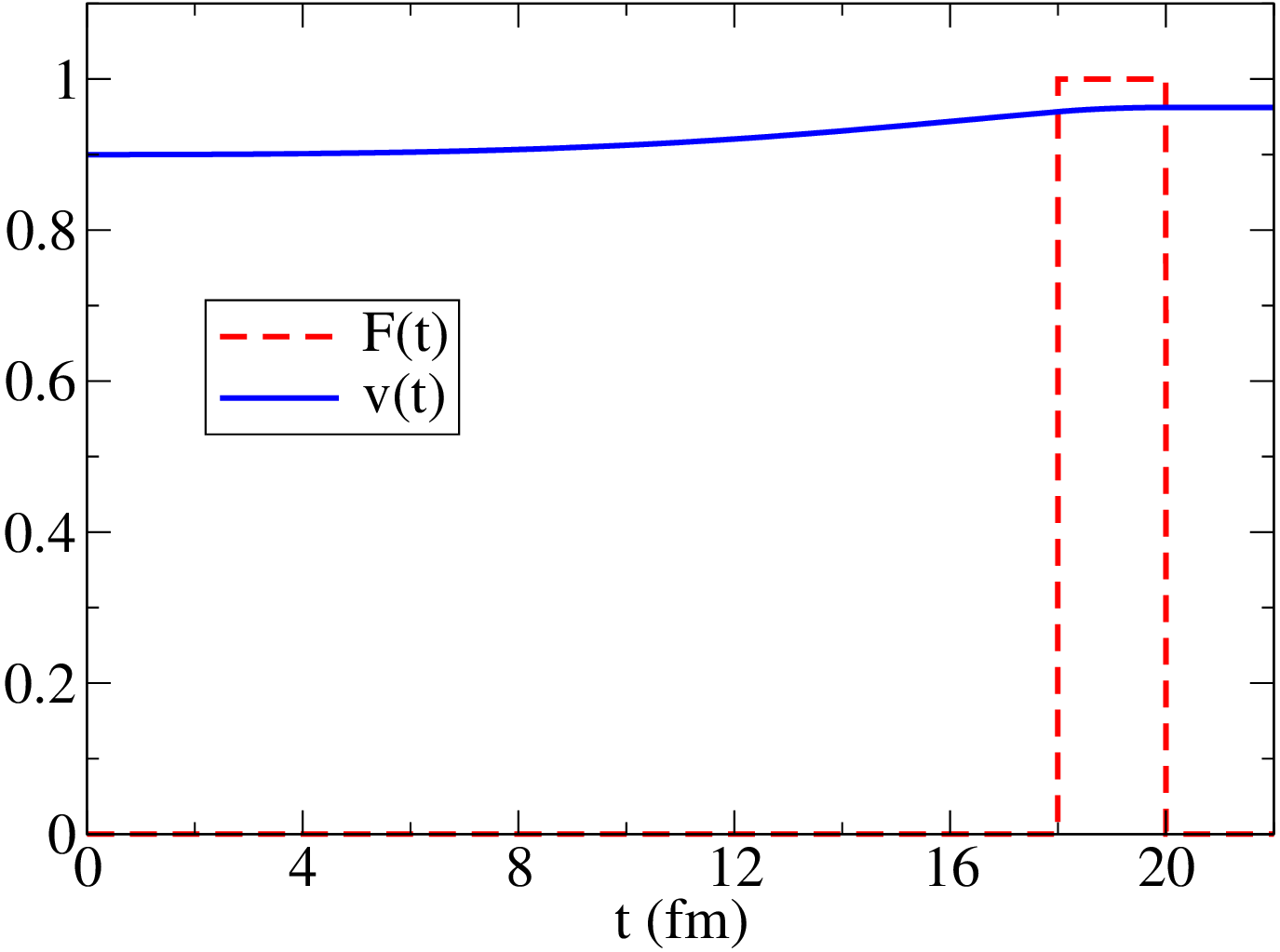}
\caption{Same as in figure~\ref{fig:RadStep1}, except the particle moves initially along the force axis with velocity $v/c=0.9$. The actual acceleration $\ar {\bf a}\ar $ (dotted line in the top plot, note the log scale) is thus much less pronounced. In addition we also plot the velocity (solid line in the bottom figure). (Units as in figure~\ref{fig:RadStepFlat}).}
\label{fig:RadStep2}
\end{figure}

In figure~\ref{fig:RadStep2} the force is the same as in figure~\ref{fig:RadStep1}, 
but we set as initial condition a particle propagating (parallel to the force) with initial velocity 0.9$c$, so the effective change of the velocity (shown in the bottom plot) is much smaller.

\begin{figure}
\includegraphics*[width=8.5cm]{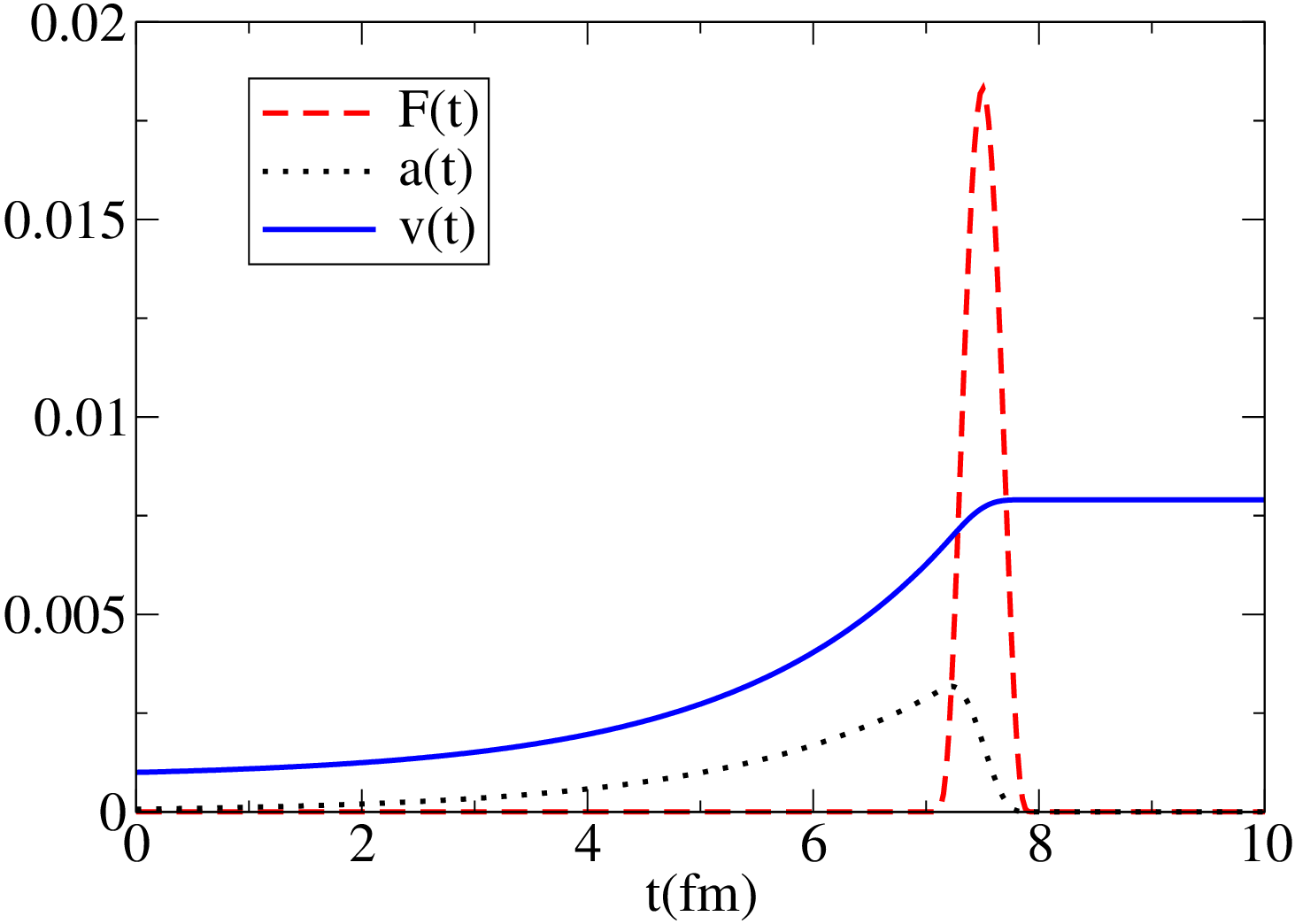}
\caption{
We soften the step-function force to make it continuous, and with all derivatives also continuous, by employing Eq.\eref{expinv}.   (Units as in figure~\ref{fig:RadStepFlat}).
}
\label{fig:gaussian}
\end{figure}

We now address the objection of Valentini~\cite{Valentini} discussed in subsection~\ref{subsec:causality}, stating that the Lorentz-Dirac equation was initially deduced under the hypothesis that further derivatives of the acceleration existed, and thus one should not employ a discontinuous force. We show in figure~\ref{fig:gaussian} an example where the force is proportional to a function
\be \label{expinv}
F(t) = \left\{ 
\begin{array}{ll}
0\ {\rm if} &            t<t_0 \\
e^{-(1\ {\rm fm}^2)/((t-t_0)(t_1-t))}& t\in (t_0,t_1) \\
0\ {\rm if} &            t> t_1 
\end{array}
\right.
\ee
that, although not being analytic, is continuous and has continuous derivatives to any order throughout the entire $t$ real line. In this extreme (very large force acting during a very short interval) the resulting acceleration and velocity (we now take $v_0=0.001$) are similar to figure~\ref{fig:RadStep1} demonstrating that the discontinuity of the force plays no particular role: preacceleration is also present.

\begin{figure}
\includegraphics*[width=8.5cm]{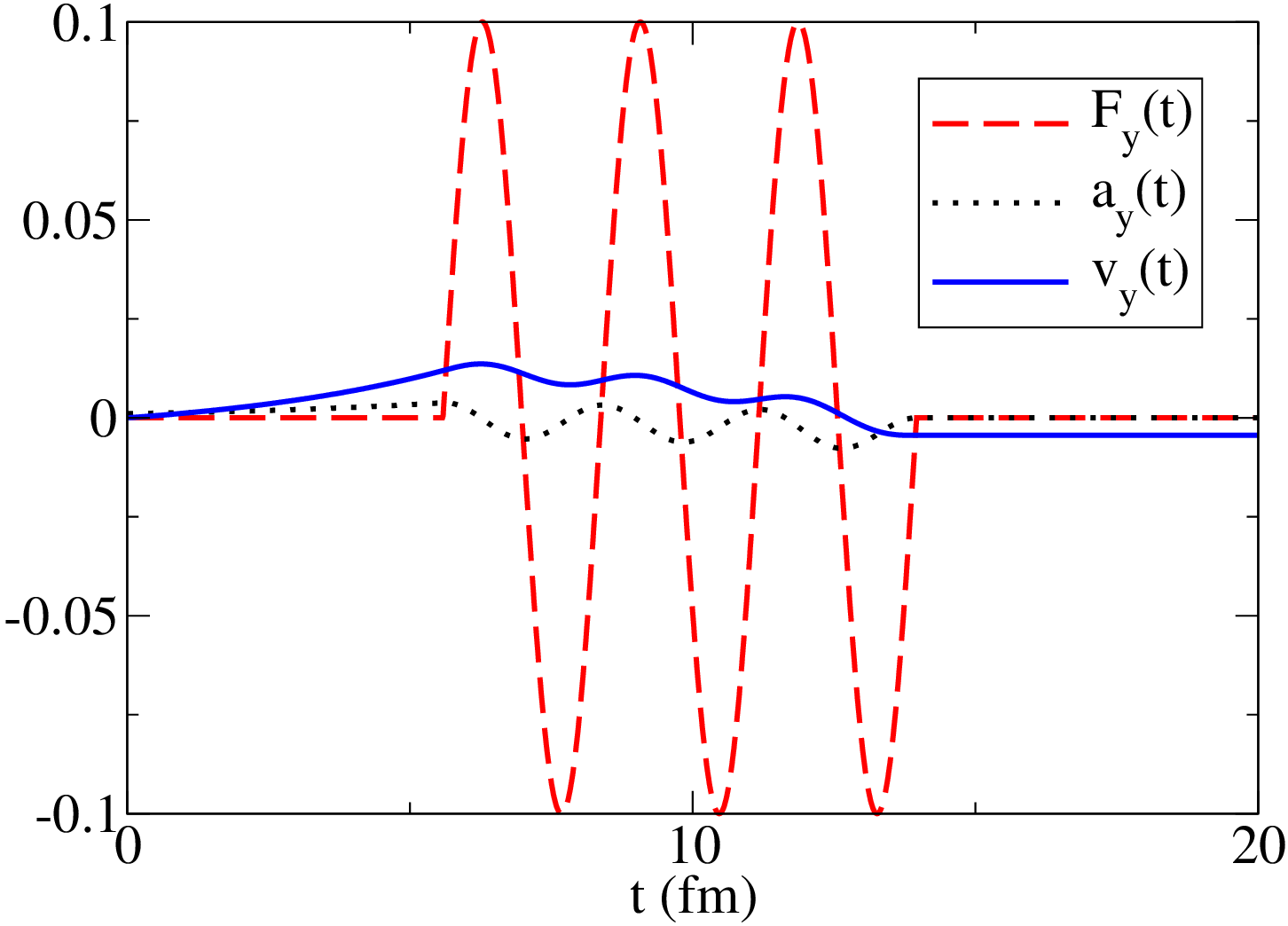}
\caption{Solution to the equation of motion under a rapidly oscillating sinusoidal force of finite extent. The phenomenon of preacceleration and the asymptotic condition of vanishing acceleration at $t\to\infty$ are clearly visible.
While the force is acting, the acceleration is dephased by about $\pi/2$ with respect to it, and they do not vanish simultaneously. 
(Units of length are fm).}
\label{fig:periodicwell}
\end{figure}

In figure~\ref{fig:periodicwell} we solve the same equation\eref{NewRadRecEq} with a periodic force that is switched on and off (modulated) by a step function, so that only three periods of the sinusoidal force are active. The solution presents pre-acceleration.

The reader should remember that in the figures we plot the laboratory force $F^i\equiv f^i_{ext}/\gamma$, that is not directly what enters the four-force in the equation of motion due to the extracted Lorentz $\gamma(v)$ factor, so that the effective acceleration is not sinusoidal,  but depends on the velocity.

%


\begin{figure}
\includegraphics*[width=8.5cm]{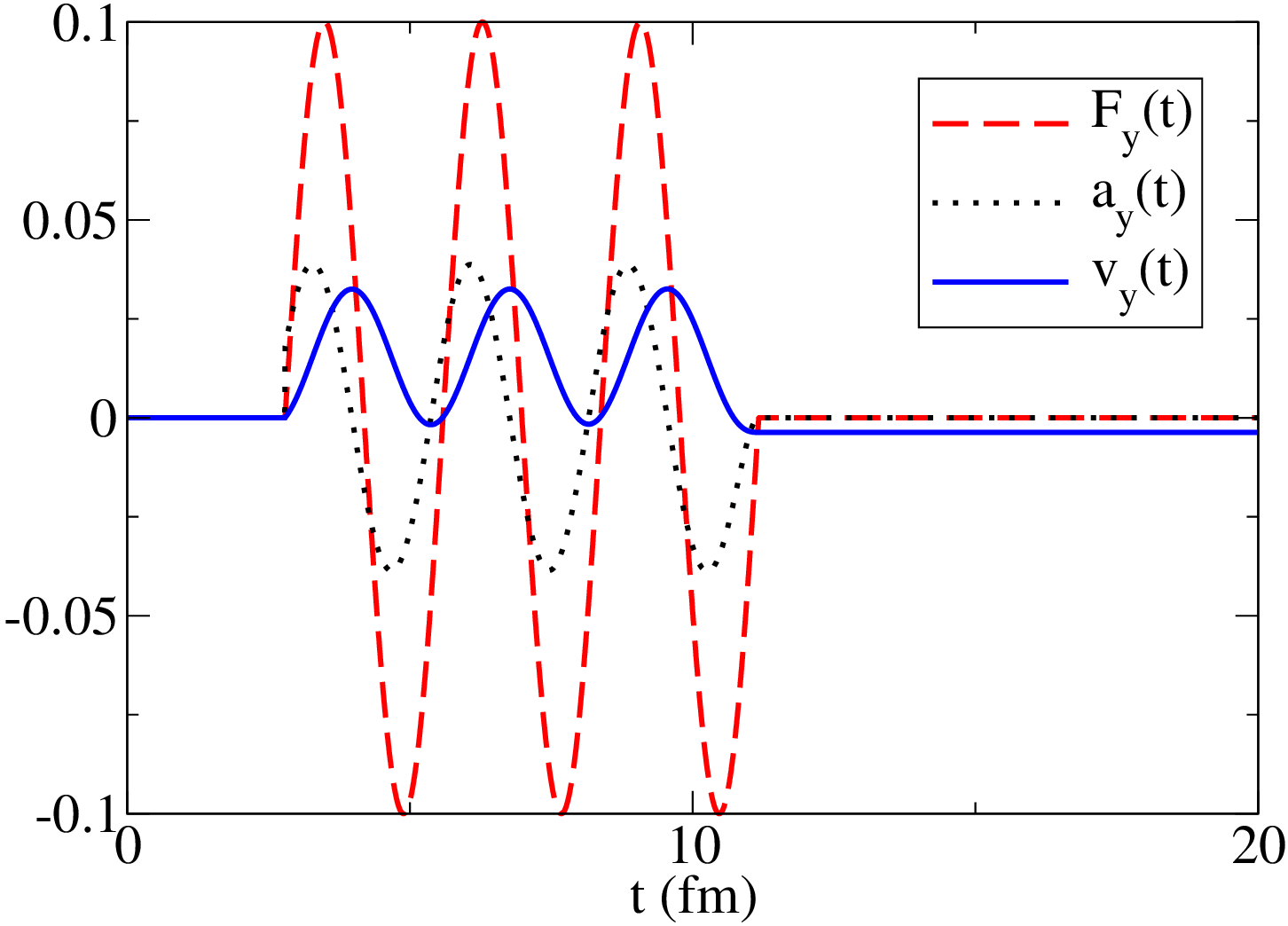}
\caption{Same as figure~\ref{fig:periodicwell} but artificially reducing the effect of radiation reaction by taking a much smaller $L\simeq O(0.1)$fm instead of the actual value $L=1.876$fm equivalent to $2\alpha/(3m_e)$. As can be seen, now the acceleration is practically in phase with the force, and in agreement with Newton's equation for the motion in an electric field without radiation reaction, showing that the integral formulation and computer code are not responsible for the preacceleration in figure~\ref{fig:periodicwell}, the cause is rather intrinsic to the theory starting with Eq.\eref{LorentzDirac}. 
Note also that the acceleration in that figure was damped with respect to this, 
as the radiation reaction much reduces the acceleration of the particle, the forces being equal. (Units of length are fm).}
\label{fig:periodicwell3}
\end{figure}

The last of this series of exercises is reported in figure~\ref{fig:periodicwell3} where the electron relaxation length $L$ is artificially diminished by a factor of about 20 to approach the non-relativistic limit of subsection~\ref{subsec:largemass}.  The acceleration is now almost in phase with the force as expected (although at zero crossing of the force it has appreciable non-vanishing values). 

\subsection{Solving the equation with radiation reaction for large
systems} \label{subsec:solvelarger}

In this subsection our aim is to demonstrate that the equation~\eref{NewRadRecEq} including radiation reaction, in spite of having a natural scale of $L=1.876$ fm for an electron, can be scaled to work with large systems. We will climb to six orders of magnitude larger lengths in laboratory time, to the nanometer scale.

The external force in the next example will be caused here by an external electromagnetic field, thus
\be
f^\mu_{ext} = e F^{\mu\nu}_{ext}(x)\dx_\nu \ .
\ee

Our first example will again be an oscillating field, but its magnitude and physical size is loosely inspired by the relatively new concept of crystal undulator~\cite{Bellucci:2003bg}.

An undulator is a well known device employed in synchrotron light sources and free electron lasers to produce electromagnetic radiation. An alternating magnetic field with a period of a few centimeters forces electrons to wiggle and emit synchrotron light or X-rays, that are collected for applications. Our equation with radiation reaction could  be applied to these systems, but we will start here with a smaller device, for an exercise with less computing power, where in several thousand steps the effect of radiation reaction is clearly visible.

In a crystal undulator, positrons (and somewhat less effectively, because of their attraction to the positively charged crystal sites, electrons) are channeled between
crystal planes. The elementary charge can suffer a hard collision with another electron in the medium or with a nucleus. Such processes require a Quantum Electrodynamical description~\cite{Kostyuk:2011kh,Shulga} and are beyond our scope. However, the continuous interaction with the crystal oscillating Coulomb field resulting in the emission of relatively low-frequency radiation (as opposed to hard photons) is amenable to a classical description.

\begin{figure}
\includegraphics*[width=8.5cm]{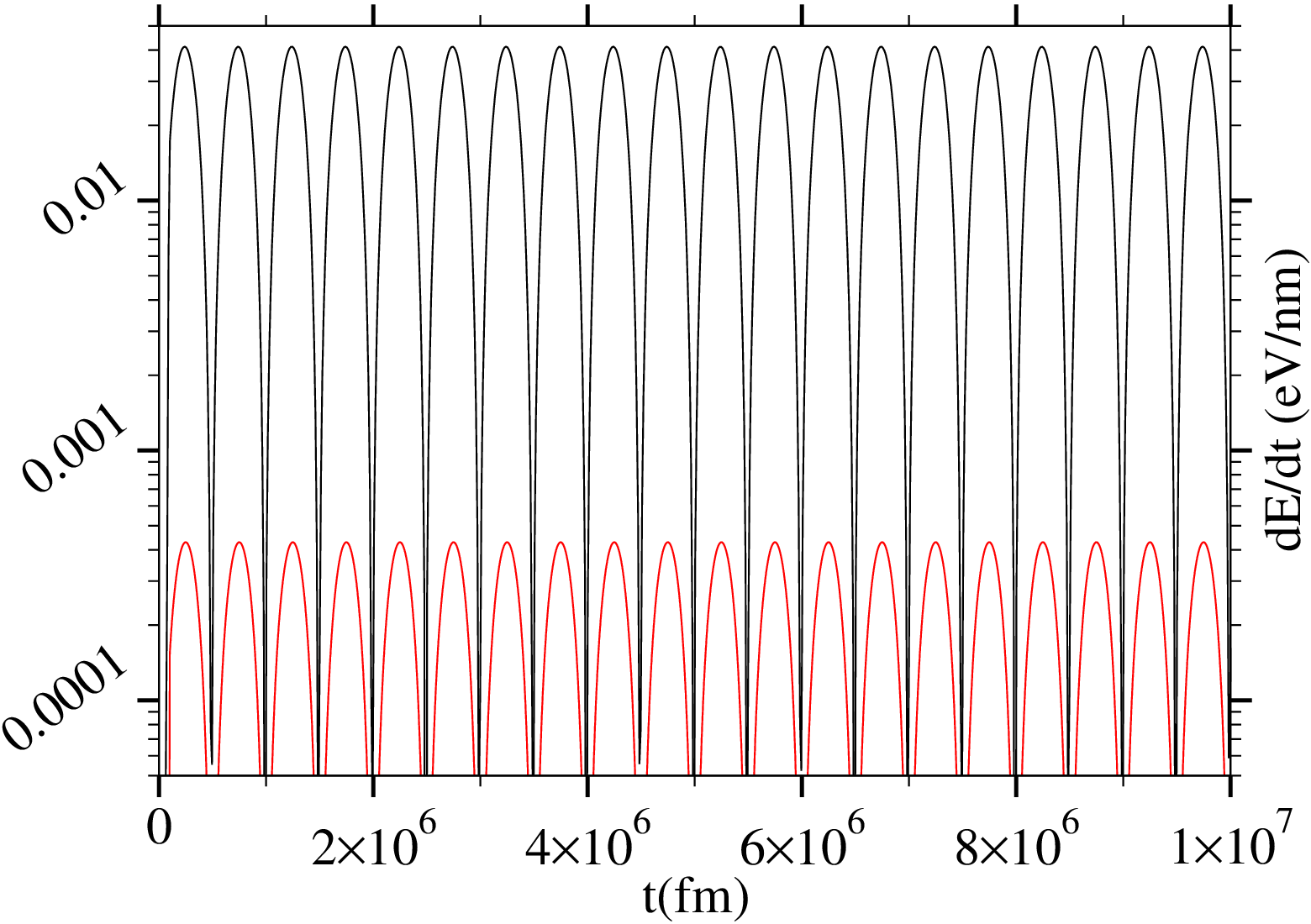}
\caption{Power radiated by an electron channeled along a toy crystal of physical nm dimension for energy 300 MeV (lower curve, red online) and 3 GeV (top curve, black online). The full Eq.~(\ref{NewRadRecEq}) is used.
\label{fig:undulator1}}
\end{figure}

We take as a model electric field ${\bf E} =E_0 \sin (2\pi x/a) {\bf e}_{y}$ with $E_0\simeq e/r^2$, and a typical nanometer length $r\simeq a = 1$ $10^6$ fm. Thus the electron, with initial velocity $v_0 {\bf e}_{x}$, perceives a perpendicular oscillating field. 
One can characterize this field by a length scale $\Lambda_E \equiv \frac{m_e}{eE_0}$ that is of order $2.8\times 10^{10}$ fm. Thus, there is a hierarchy of scales
$L\ll a \ll \Lambda_E$ and we are happy that the computer code can handle all three scales without special attention. \\
Taking $L$ as the scale of the computation, $\Lambda_E$ is a huge number, so that the force $\propto \frac{1}{\Lambda_E}$ is tiny, but if left to act for large times above $a$, it certainly has an important influence on the trajectory.

The trajectory is not particularly interesting, but we follow the emission of radiation for electrons of two energies, 300 MeV and 3 GeV respectively, in figure~\ref{fig:undulator1}.

\begin{figure}
\includegraphics*[width=8.5cm]{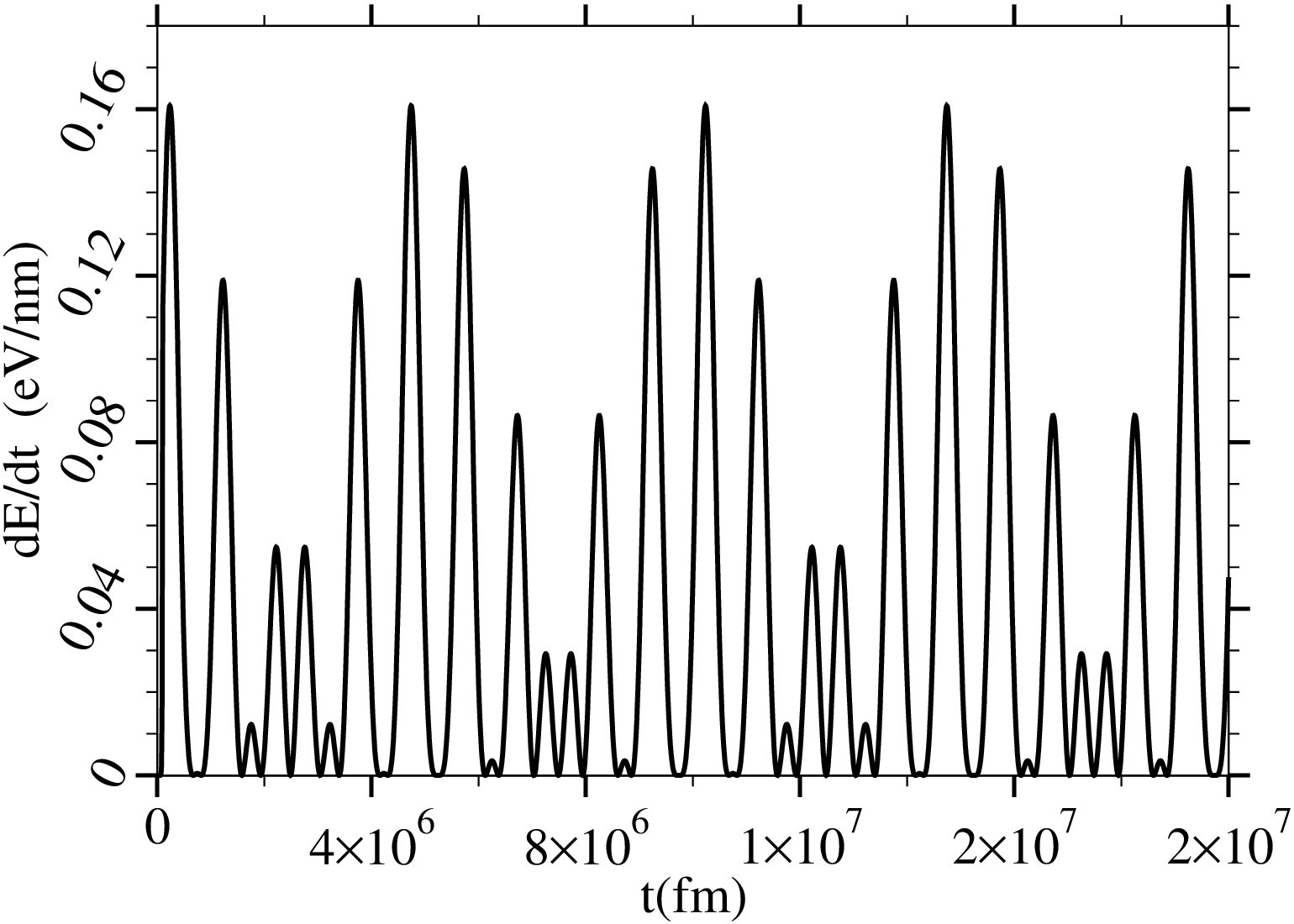}
\caption{Power radiated by a 3 GeV electron channeled along a toy crystal with initial velocity along the $OX$ axis. An oscillating field along $OY$ has period 1 nm. An additional field rotating in the $XY$ plane, and of equal strength than the first,  has period 10 nm (simulating a wavy crystal deformation);
the periodicity of the motion can be appreciated.
The total energy radiated is here much smaller than the total energy.
\label{fig:undulator2}}
\end{figure}

 To make the example more interesting, to this electric field we add an additional field of equal intensity but rotating in the following way, 
$$
{\bf E'}=E_0 \left( \sin(2\pi x/(10a)){\bf e}_{x}+ \cos(2\pi x/(10a)){\bf e}_{y}\right)\ ,
$$
that is, with a period ten times larger. This doesn't have any claim of realism to describe an existing device, but again it is just used to demonstrate the capability of our equation. For real undulators the period should at least be a factor of 10 larger, approaching even the micrometer scale, but maybe our calculation is more directly applicable to so called ``nanowigglers''~\cite{Batrakhov} that are conceived at the nanometer scale. 

At any rate, the motion of the electron emitting classical radiation can be numerically followed, as we do in fig.~\ref{fig:undulator2}, concentrating again on the emission of radiation, that presents strong oscillations and interference effects between the two periodic fields.

We now turn to our second example, inspired by the physics near a neutron star crust. 
We take a positron (emitted in beta decay or by any other means) with an initial Lorentz factor of $\gamma=10$ to be injected in a constant and homogeneous magnetic field along the $OZ$ axis of intensity $B_0=10^{14}$ Gauss, at the upper limit of the conditions usually agreed for such stars.
The characteristic length for the field intensity is $\Lambda_B = \frac{m}{eB_0}\simeq 470$ fm, so that the hierarchy of scales is not as marked as in the previous example.

\begin{figure}
\includegraphics*[width=8.5cm]{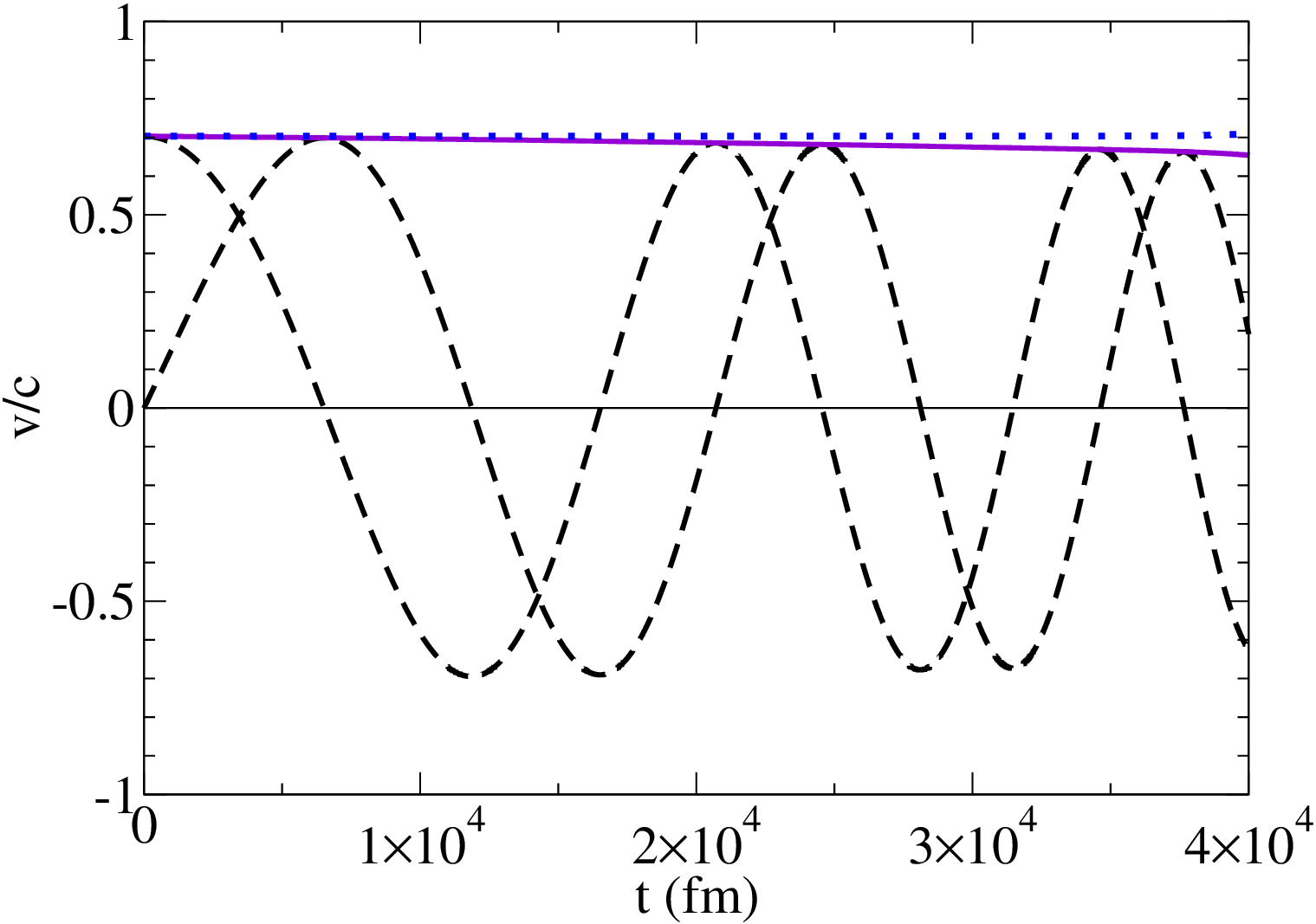}
\caption{A positron with energy 5.1 MeV and initial velocity along $\frac{1}{\sqrt{2}}(1,0,1)$ is injected into a region with a constant, homogeneous magnetic field ${\bf B}= B
{\bf e}_{z}$. We plot $v_x$ and $v_y$ (dashed lines, $v_y$ starting at zero), shifted by $\pi/2$, as well as the total $v_\perp$ (solid line) and $v_z$ (constant dotted line).
\label{fig:vB}}
\end{figure}

\begin{figure}
\includegraphics*[width=8.5cm]{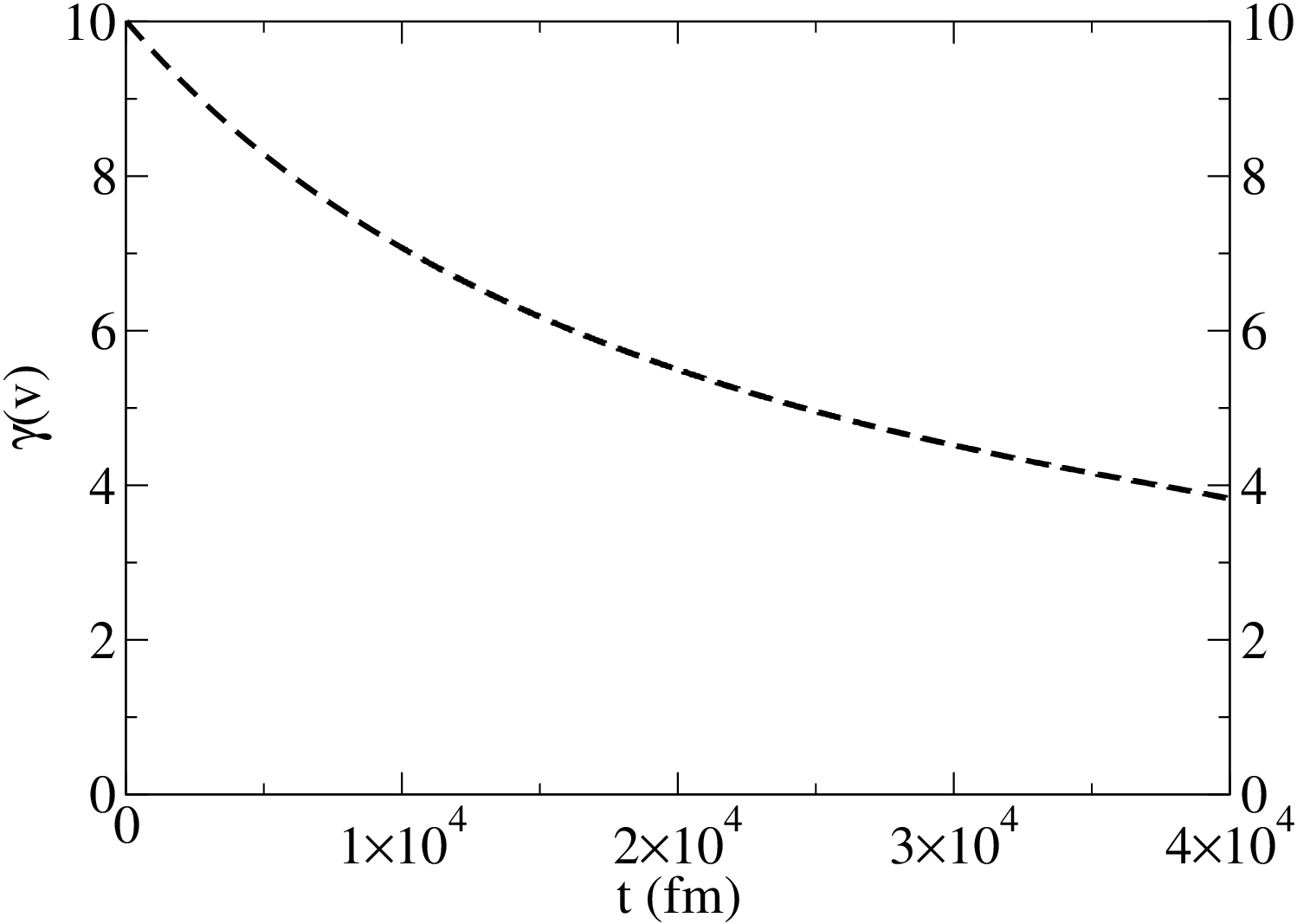}
\includegraphics*[width=8.5cm]{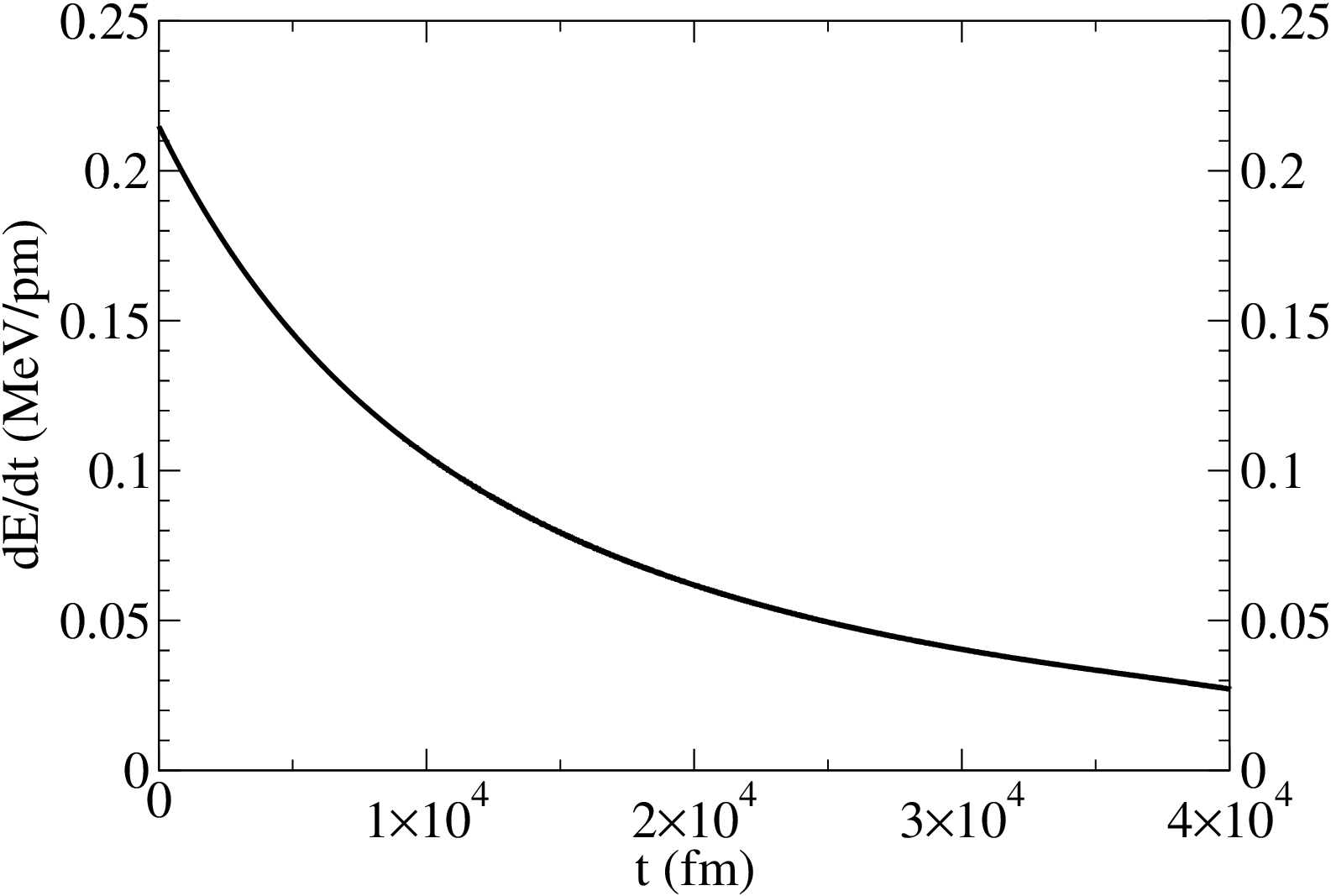}
\caption{ For the same particle and field as in figure \ref{fig:vB},
we plot the Lorentz contraction factor, or kinetic energy per unit mass (top plot). 
Since a magnetic field does not do work on a charged particle, the energy lost must be carried away by the radiation.
 The computation of ${\mathcal R}$ is shown in the bottom plot.
\label{fig:EB}}
\end{figure}

Neglecting first the radiation reaction, the trajectory is helicoidal around the magnetic field ${\bf B} = B {\bf e}_{z}$, that projects to a circumference on the $XY$ plane with radius $r_\perp = 4.7\times 10^3$fm.  This trajectory provides a good zeroth order guess $x_{[0]}$ for the program iteration. Of course, the energy and gyration radius will vary once radiation reaction is included.

Figure~\ref{fig:vB} shows the velocity corresponding to having solved Eq.\eref{NewRadRecEq} for this field. Initially we inject the positron with equal 
$$
v_z=v_\perp\equiv\sqrt{v_x^2+v_y^2}
$$ 
at the origin of coordinates (therefore the axis of the helicoidal motion is parallel to, but not coincident with the $OZ$ axis). However, the emission of radiation slows down the electron so that $v_\perp$ decreases as seen in the figure (the maximum $v_x$ and $v_y$ values are decreasing).

Although the total velocity remains above 0.9$c$ (having started with 0.99$c$), we see in figure~\ref{fig:EB} that the energy per unit mass drops rapidly to less than half its initial value in only three oscillations, below the nanometer scale. 
This is natural since the radiated synchrotron power for a circular trajectory (taken from Eq.(5.14) in Eq.~\ref{Rohrlich}, up to the metric's signature)
from 
\be \label{radpower}
\mathcal{R} = 
-\frac{2\alpha}{3}  \ddx^2
\ee
grows with the fourth power of the energy, 
\be
{\mathcal R} = 
\frac {2\alpha}{3} \gamma^4 \left( \frac{v^2}{r} \right)^2\ ,
\ee 
so the trajectory will be much more severely distorted for large velocities.
The loss of energy is compensated by this emitted radiation (bottom plot of figure~\ref{fig:EB}) since the magnetic field performs no work on the particle.
\\
The asymptotic condition in this calculation is implemented here simply by the fact that, as the particle radiates energy away, the Lorentz force $e \frac{\bf v}{c}\times {\bf B}$ vanishes with the velocity and the particle reduces its radius of gyration and is thereafter left to coast freely in the $z$ direction with no external acceleration. So at some large $s_{max}$ we simply set the acceleration to zero.

\begin{figure}[h]
\includegraphics*[width=6.5cm]{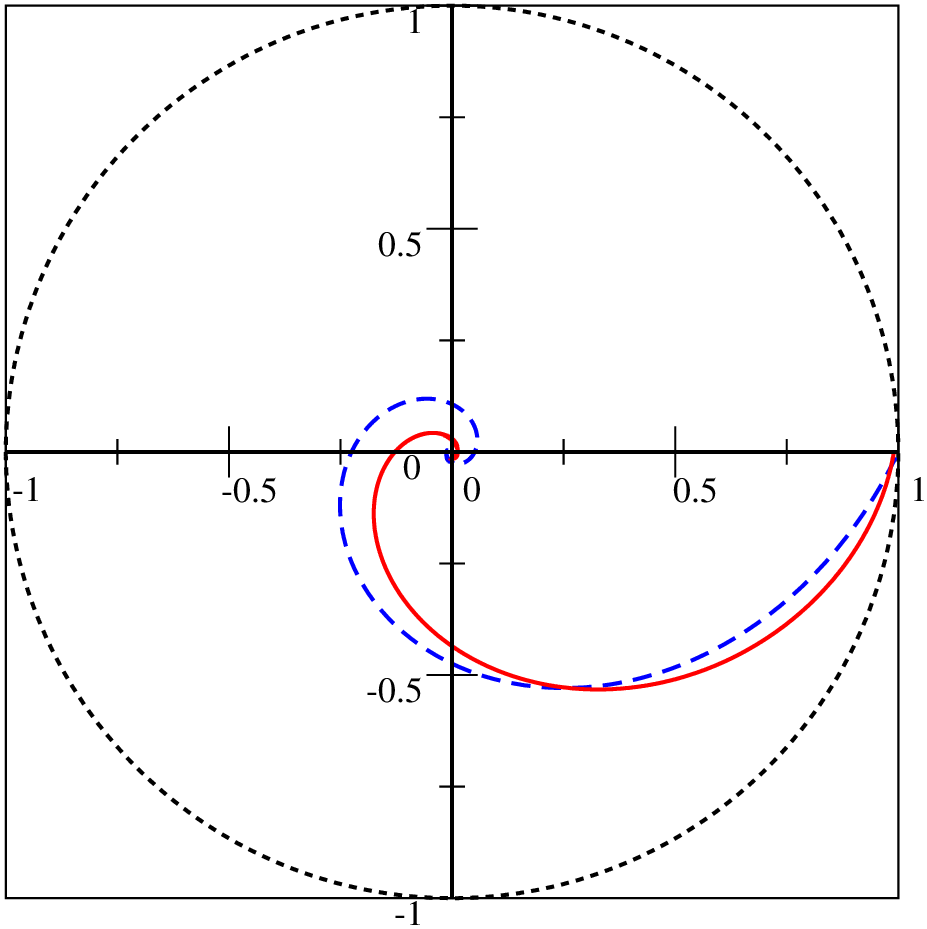}
\caption{ Velocity-space trajectory in the $(v_x,v_y)$ plane for a particle initially moving along the $OX$ axis inside a magnetic field $B\hat{z}$
\label{fig:Plass}
Solid red line: our computation. Broken blue line: Plass's non-relativistic approximation in figure 9 of~\cite{Plass:1961zz}. Dotted outer circle: motion without radiation reaction.  Note that neither trajectory  reaches the origin for finite time; if we zoom towards the origen, we see that the electron continues inspiralling indefinitely.}
\end{figure}

A closely related example has been considered by Plass~\cite{Plass:1961zz}. Plass, due to computer limitations, employs a non-relativistic approximation to a charged electron in a constant and uniform magnetic field that is way stronger than our example (already a very large ${\bf B}$ at a neutron star). The trajectory in velocity space, projected on the plane $(v_x,v_y)$, is depicted in figure~\ref{fig:Plass} and it closes towards the origin very fast (very strongly damped motion). 
 We have taken the $v_z$ velocity component as vanishing at $t=0$ (and thus, at all times), as well as a dilatation factor $\gamma=10$ as initial condition. As the initial velocity is thus 0.99, it is not surprising that our fully relativistic calculation differs appreciably from Plass. 

\begin{figure}
\includegraphics*[width=6.5cm]{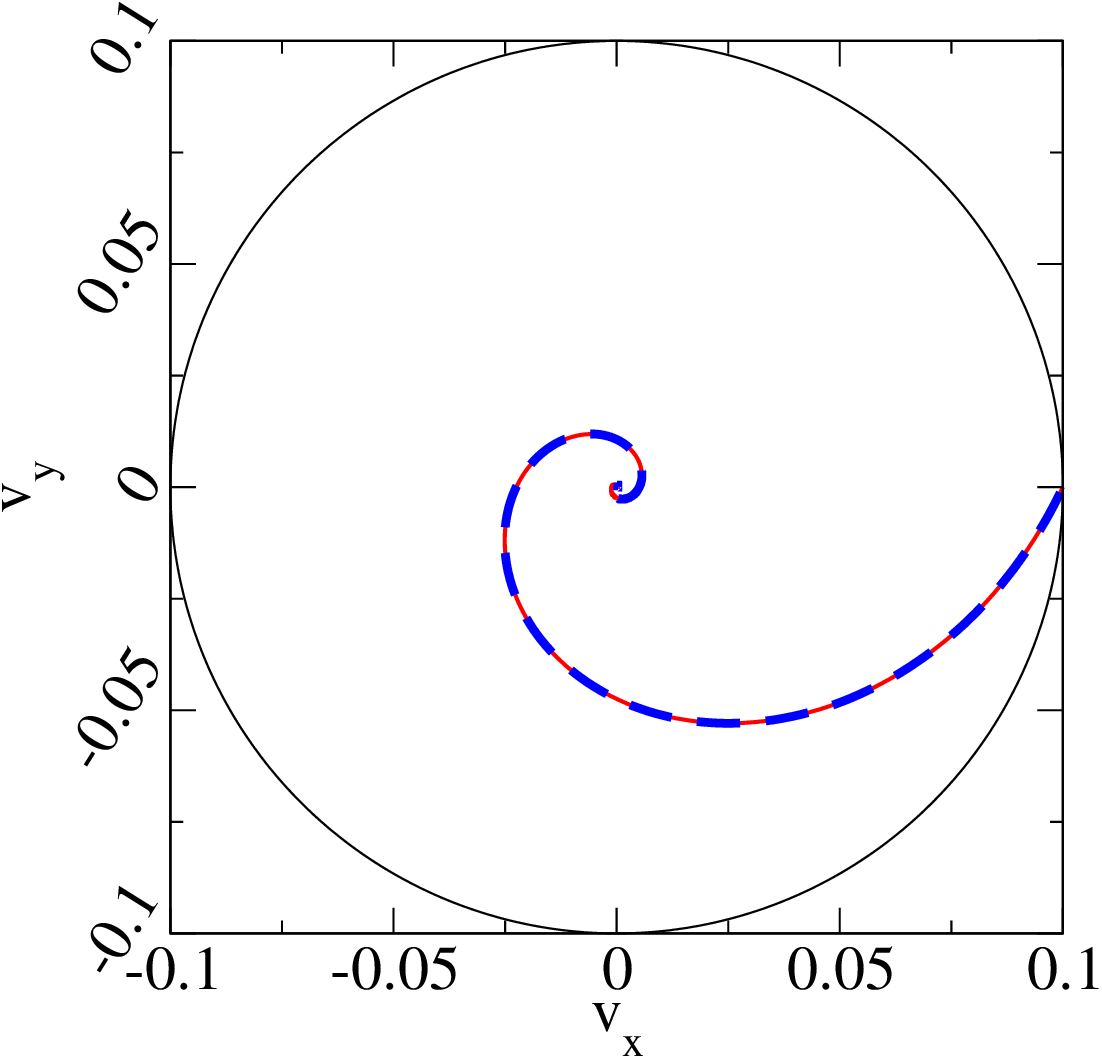}
\caption{Velocity-space trajectory in the $(v_x,v_y)$ plane for a particle initially moving along the $OX$ axis inside a magnetic field $B\hat{z}$; the velocity here is taken to be 0.1 ($c=1$).
\label{fig:Plass2}  Solid red line: our computation. Broken blue line: Plass's non-relativistic approximation.  The agreement is now excellent. The distance of the electron to the origen vanishes exponentially, without reaching it at any finite time.}
\end{figure}

To check agreement with Plass we have also considered an initial condition with the same strong magnetic field but an initial velocity $v=0.1$, where the approximation of Plass should work much better. We find that our numerical computation falls exactly on top of the analytical result (see figure~\ref{fig:Plass2}).

\begin{figure}
\includegraphics*[width=6.5cm]{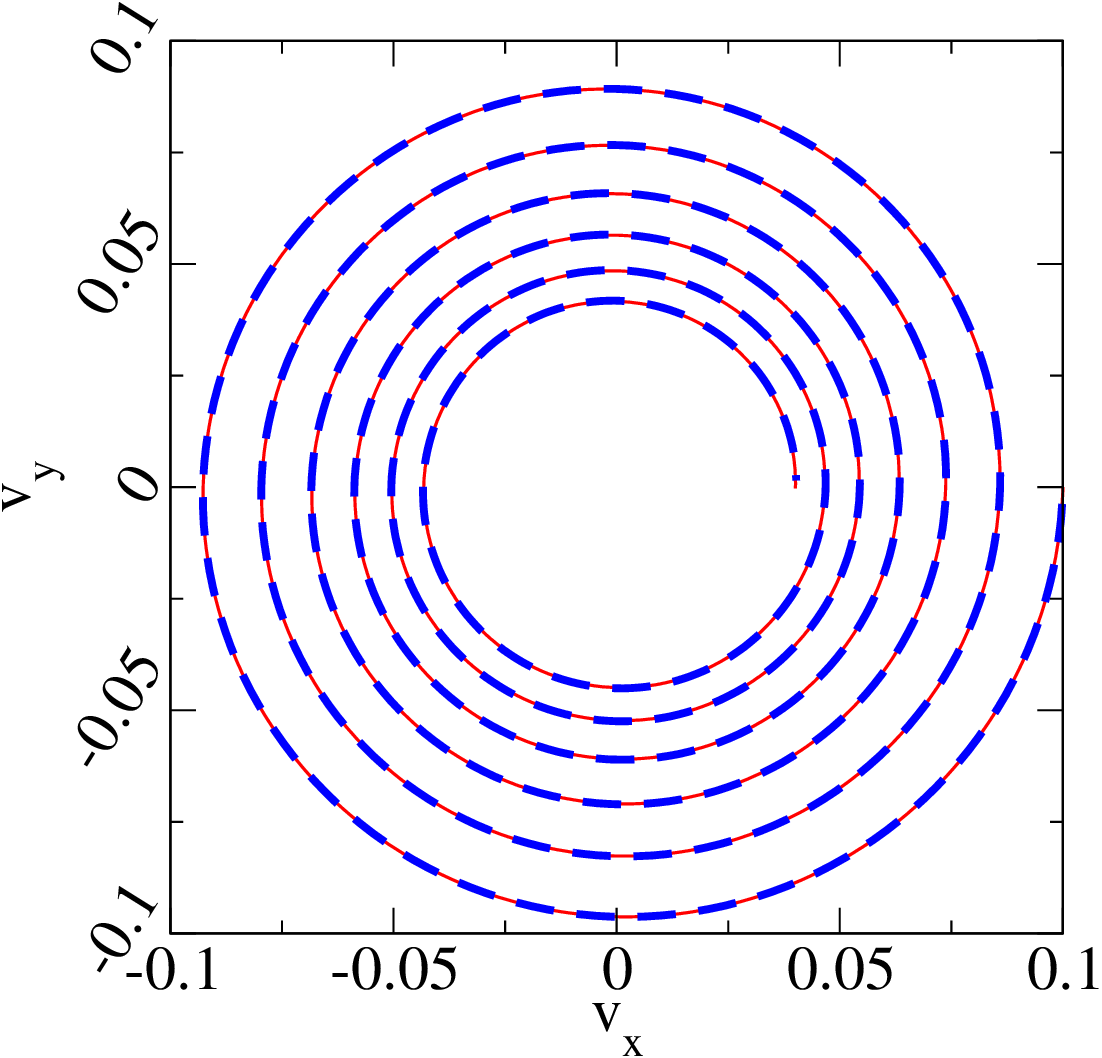}
\caption{ Velocity-space trajectory in the $(v_x,v_y)$ plane for an electron with the initial condition $v=0.1$ and field 2.5\% of figure~\ref{fig:Plass2}. Again, the agreement is excellent. The distance betweeen consecutive orbits decreases as the electron slows down.
\label{fig:Plass3}}
\end{figure}
Finally we have considered an initial velocity $v_0=0.1$ but a weaker magnetic field. The agreement with Plass is again perfect (within our numerical precision of two parts in ten thousand).

The dynamical situation with both initial velocity and magnetic field taking large values, depicted in figure~\ref{fig:Plass}, is numerically somewhat more subtle due to poor stability of the solution at intermediate steps of the numerical computation. We have improved on the algorithms described so far incorporating a Runge-Kutta integrator for the differential equation and a piecewise integration dividing the total proper time interval. The results shown are converged to better than a per thousand precision.

Thus, we have shown in this subsection the possibilities that the new formulation opens to accommodate non-trivial calculations of physical interest.

\section{Summary and discussion}\label{sec:conclusions}

In this work we have presented a modified equation of motion for a classical particle under the influence of radiation reaction. We have not attempted to solve the problem of preaceleration, which is a 
pathology associated with the Dirac-Rörlich integro-differential equation. (Differential-equation treatments for point particles, such as the Lorentz-Dirac or Landau-Lifschitz formulations, do not present it, and of course we do not presume that it is a characteristic of Nature either). 
This unpleasant feature could be avoided considering extended charge distributions, but this is beyond the scope of this paper. 

Nevertheless, although our alternative equation\eref{NewRadRecEq}  does not cure this flaw, it is superior in several technical advantages for its numerical implementation that may make it useful for systems as varied as the astrophysics of neutron stars, or the synchrotron light emitted by a crystal undulator or nanowiggler.
We have by far not exhausted the number of areas where equation\eref{NewRadRecEq} could be of use. For example, intense laser fields are increasingly becoming available and awakening interest in the community~\cite{Piazza}, and reaching $10^{22}$ watt per square centimeter~\cite{Seto}, they are believed to enter the regime where radiation reaction is important. Since the field is characterized by extreme oscillations, our worked examples could be extended to cover the motion of a charge inside such field.

Another striking example where we should like to apply our equation in the future is ``classical tunnelling'', 
possibly a consequence of preacceleration, by which an electric charge can pre-increase its energy of motion 
by absorbing it from the  near field, to shed it later, thus passing classically above nominally forbidden barriers. This concept has been demonstrated for the Lorentz-Dirac equation~\cite{Denef:1996ng} and we are curious about whether the integral formulation here presented also supports it. This interesting feature deserves a complete  (and thus, labour intensive) analysis that we postpone for another work.
Our treatment of this problem by means of the integral formulation will avoid the conundrum of separating self-accelerated solutions in the presence of a force, and will thus have different systematic uncertainties.

We have found, confirming the work of other authors~\cite{Denef:1996ng}, that discontinuities of the force function or its derivatives are a feature that is not essential for the counterintuitive behavior of the motion influenced by radiation reaction.

The differential equation of Landau-Lifschitz shares with our integral version of the equation of motion the convenient feature of quasilinearity, the highest (second) derivative of the position being isolated on the left-hand side as in a classical Newtonian formulation. Indeed, in section IV C we have also derived a relativistic version of the Landau-Lifshitz equation as an approximation to the integral equation of motion.



The new equation\eref{NewRadRecEq} (or new formulation of the existing equation, if one wishes)
is linear in the highest (second) derivative, that is explicitly solved for and not left implicit. Practitionners of numerical analysis in classical electrodynamics will no doubt find this and the other advantages of the new equation superior and useful. We look forward to the further applications of the theory.


\begin{acknowledgments}
The authors would like to thank Norbert M. Nemes and Juan Ramirez Mittelbrunn for reading the manuscript.
We also thank the constructive comments and additional references provided by the anonymous referees. This work was supported by spanish grants FPA2011-27853-01, FIS2008-01323 and UCM-BS GICC 910758.  Felipe J. Llanes-Estrada should like to dedicate this work to the memory of Maria Luz Llanes Men\'endez.
\end{acknowledgments}

\end{document}